\documentclass{emulateapj}
\usepackage{multirow}
\usepackage[colorlinks,urlcolor=blue,citecolor=blue,linkcolor=blue]{hyper ref}
\usepackage{graphicx}
\usepackage{amsmath}
\usepackage{threeparttable}

\begin{document}
\title{Chemical network reduction in protoplanetary disks}

\author{Rui Xu\altaffilmark{1}, Xue-Ning Bai\altaffilmark{2,3}, Karin \"{O}berg\altaffilmark{4}, Hao Zhang\altaffilmark{5}}
\altaffiltext{1}{Department of Astrophysical Sciences, Princeton University, Princetion, NJ 08544;
ruix@princeton.edu}
\altaffiltext{2}{Institute for Advanced Study, Tsinghua University, Beijing 100084, China; xbai@tsinghua.edu.cn}
\altaffiltext{3}{Tsinghua Center for Astrophysics, Tsinghua University, Beijing 100084, China}
\altaffiltext{4}{Harvard-Smithsonian Center for Astrophysics, 60 Garden St., Cambridge, MA, 02138}
\altaffiltext{5}{Department of Physics, Purdue University, 525 Northwestern Ave., West Lafayette, IN 47907}

\begin{abstract}

Protoplanetary disks (PPDs) are characterized by different kinds of gas dynamics and chemistry, which are coupled via ionization, heating and cooling processes, as well as advective and turbulent transport. However, directly coupling gas dynamics with time-dependent chemistry is prohibitively computationally expensive when using comprehensive chemical reaction networks. In this 
paper, we evaluate the utility of a species-based network reduction method in different disk environments to produce small chemical networks that reproduce the abundances of major species
found in large gas-phase chemistry networks. We find that the method works very well in disk midplane and surfaces regions, where approximately 20-30 gas phase species, connected by $\sim$50-60 gas phase reactions, are sufficient to reproduce the targeted ionization fraction and chemical abundances. Most species of the reduced networks, including major carriers of oxygen, carbon and nitrogen, also have similar abundances in the reduced and complete network models. Our results may serve as an initial effort for future hydrodynamic/magnetohydrodynamic simulations of PPDs incorporating time-dependent chemistry in appropriate regions. Accurately modeling the abundances of major species at intermediate disk heights, however, will require much more extended network incorporating gas-grain chemistry and are left for future studies.

\end{abstract}

\section{Introduction}

The formation of planets, and perhaps the existence of  extraterrestrial life, is intrinsically linked to the chemistry of planet formation. Chemical processes in protoplanetary disks (PPDs) offer rich observational diagnostics that help constrain disk properties (e.g. \citealp{WilliamsCieza11,HenningSemenov13}). 
The volatile content in PPDs is directly relevant to the abundance and composition of solids \citep{Oberg2011}, which affects grain surface properties
and dust coagulation \citep{Wada_etal_09,Okuzumi_etal_16}, and
will eventually be reflected in the composition of planets \citep{Madhusudhan11}. Formation of complex organic molecules in disks may serve as initial seeds of extraterrestrial life \citep{Oberg_etal_15}. 

 Chemical processes in PPDs are coupled with many aspects of the PPD gas dynamics. The level of disk ionization (including charge distribution), resulting from ionization-recombination chemistry, largely determines the coupling between gas and magnetic field (e.g. \citealp{UmebayashiNakano1988,Sano2000,Fromang_etal_2002,Ilgner06,Okuzumi2009,Aikawa_etal_2015}). Such coupling further governs the overall gas dynamics and angular momentum transport, and eventually determines the global disk structure and evolution (e.g., \citealp{Wardle07,Turner_etal14,Xu_Bai_16,Bai_17}). In the disk atmosphere, heating and cooling processes are also directly coupled with chemistry \citep{Glassgold04,Nomura05,Kamp_etal_2010}, which determines the thermal structure in the disk surface layer. The heating and cooling processes also strongly affect disk mass loss through (magneto-)photoevaporation \citep{Gorti_etal_09,Owen_etal_10,Bai_etal16,Bai16}, and contribute to disk dispersal. Moreover, detailed knowledge of thermodynamics and chemistry is essential in interpreting observations of emission lines from the warm disk surface layer \citep{Carr_etal_04,Simon_etal_16}. More recently, we showed in \citet{XuBaiOberg17} the impact that gas dynamics can have on CO depletion in the upper disk atmospheres.

Despite the evidence above that chemistry and gas dynamics in PPDs affect each other, they are generally modeled independently. Most studies of disk gas dynamics assume a steady-state chemistry. For instance, tabulated magnetic diffusivities based on steady-state ionization chemistry are commonly used in non-ideal magnetohydrodynamic (MHD) simulations of PPDs (e.g., \citealp{BaiStone13,Gressel_etal15,Bai_17}). Disk chemistry studies have revealed that in most disk regions steady-state is never reached, however, but rather the chemistry continues to evolve time dependently throughout the disk life time.

It is highly desirable to couple realistic gas dynamics and chemistry in time-dependent simulations of PPDs. In recent years, the growing computational power has encouraged more efforts towards this direction. Examples include star-formation in molecular clouds \citep{Glover_etal10}, and on the simulation of disk ionization (e.g. \citealp{Nomura_etal_09,SemenovWiebe11,Heinzeller_etal_11, Furuya_etal_2013,Furuya_etal_2014})\footnote{There are other works focusing on the chemistry from core collapse to early stages of disks that follow the chemistry along either analytical trajectories or post-processing of simulation trajectories without feedback \citep{Visser_etal_2009,Visser_etal_2011,Yoneda_etal_2016}.} and photoevaporation (e.g. \citealp{Wang_etal_17,Nakatani_etal_17}).  However, the stiffness of chemical reaction networks requires implicit solvers that constantly invert matrices, where the cost scales as the number of species cubed. Therefore, evolving (magneto-)hydrodynamics with time-dependent chemistry is prohibitively expensive if a large chemical network is adopted. Existing works of this kind all rely on substantially reduced chemical network to minimize computational cost, and it is the goal of this paper to test the validity of reducing network size in the context of PPDs.

Small chemical networks can be constructed by either selecting important species and reactions from large, comprehensive chemical networks \citep{Ruffle02,Wiebe03}, or by building it up from scratch using chemical intuition \citep{Keto_etal_14}. In the case of PPDs, we favor the first reductive approach since the chemistry is not always intuitive
and the latter approach therefore risk omitting key reactions and reactants. A species-based approach was originally described by \citet{Tomlin92} and \citet{Pilling97} in the context of combustion chemistry, and was later introduced to astrochemistry by \citet{Ruffle02} and \citet{Rae02}. An alternative, reaction-based reduction method was presented by \citet{Wiebe03} based on selecting important reactions rather than species. Using both methods,
\citet{semenov04} (hereafter SWH04) studied chemical network reduction for PPDs, aiming at accurately reproducing the ionization fractions. They found good agreement between ionization fractions in the reduced and full networks, but had to develop a new reduced network for each disk region (defined by a disk radius and height), which somewhat limits the utility of the presented networks.

In this paper, we adopt the species-based method, which is simpler and gives more definitive criterion in selecting species and reactions. Compared to SWH04, we expand the scope of the inquiry to also include major coolants and some abundant molecules of interest to astrochemists and planet formation theorists. We also investigate the cost of adding key photoprocesses to the disk atmosphere reduced network (e.g. self-shielding) and the cost (in terms of network size) of combining reduced networks at different disk radii to enable broader applications. Our network reduction focuses on the the disk midplane (within $\sim2$ scale heights) and disk surface layers (where FUV can penetrate). The chemistry in these two regions are relatively simple, yet has wide applications in the disk dynamics community. 

In the midplane region, the ionization-recombination chemistry can strongly affect the gas dynamics via non-ideal MHD effects (e.g., \citealp{Turner_etal14}). 
While recombination timescales in disks are typically shorter than dynamical timescales, justifying steady-state approximations \citep{Bai2011}, they become comparable in the absence of small grains. For this latter case, \citet{Turner_etal07} showed that incorporating time-dependent chemistry in a grain-free network can help revive the Ohmic dead zone in the inner PPDs in the conventional picture of layered accretion, and the free-electron abundance depends on the chemistry. Coupling gas dynamics with chemistry in this region is also important when studying advective and turbulent transport of chemical species, especially across snowlines, where dynamical timescales are often comparable to chemical timescales in the disk midplane \citep{StevensonLunine88,KretkeLin07,Krijt2016}.

In the disk surface, the mass loss process via (magneto-)photoevaporative wind is sensitive to the heating/cooling processes, which are intrinsically coupled to chemistry. In this region, chemical species can experience rapid varying environments in terms of density, temperature and radiation field due to advection in the outflow, invalidating the assumption of chemical equilibrium \citep{Wang_etal_17}. Coupling chemistry with dynamics in the disk surface layer is also crucial to provide reliable observational diagnostics to constrain disk and wind properties. 

 Chemical networks containing a few tens (rather than hundreds) of species are typically manageable in simulations where the computational cost spent on chemistry calculations is comparable or smaller than that spent on hydrodynamics/MHD (e.g., \cite{WangGoodman17}), and this is the goal of our network reduction in the disk midplane and surface layers for the above contexts. On the other hand, chemistry at the intermediate layer is known to be much more complex (SWH04), and may not be amendable to substantial network reduction.

The paper is organized as follows. In \S\ref{sec:setup}, we describe our original chemical network, disk model and the network reduction method. In \S\ref{sec:result}, we present the results of our reduction in disk midplane and disk surface, and compare it with previous works.  We summarize and discuss the applications in \S\ref{sec:con}.

\section{The Methodology of Network Reduction}\label{sec:setup}

In this section, we first describe the chemistry model and disk model that are used for our calculations in \S\ref{sec:chemmodel} and \S\ref{sec:model}. Because the physical conditions in disks vary substantially in the vertical direction, we treat the models separately at the disk midplane and surface layers (neglecting the intermediate layer). In \S\ref{sec:reduction}, we describe the reduction method. For each layer, we conduct network reduction at different radii, and then combine the results.

\subsection{Chemistry Model}\label{sec:chemmodel}

The full chemical reaction network is a standard astrochemical nework from UMIST database which includes 9 elements (H, He, C, N, O, S, Mg, Si, Fe) and 450 gas-phase species (no isotopologues and  species with more than ten carbon atoms) \citep{UMIST12}. Based on these species, 3940 gas-phase reactions are extracted from the latest version of the UMIST database. Photo-reactions only included in the disk surface models since the disk midplane region is almost completely shielded from UV photons (see last paragraph of this subsection for how photo-reactions are treated in the disk surface models). 

A single population of dust grains is included with maximum grain charge of $\pm$ 4. Grains are assumed to be spherical with fixed size a=0.1$\mu$m and density $\rho$=3g cm$^{-3}$. Dust to gas mass ratio is fixed to $f_g=10^{-4}$ to obtain the total surface area found in more realistic grain coagulation/fragmentation simulations in PPDs \citep{Birnstiel_etal2011}\footnote{Note that from the point of view of the chemistry, what matters the most is the surface area per volume element.}. The grain-related reactions included in this otherwise gas-phase chemistry model are adsorption and thermal desorption, recombination with electrons and ions, as well as neutralization reactions among charged grain species. They are given in Table 3 and 4 of \citet{Ilgner06}, with binding energies taken from the UMIST database. In this paper, we refer to ``mantle species" as counterparts of gas-phase species adsorbed onto grain surfaces. Formation of H$_2$ on dust grains is included, where we assume a constant probability $\eta$ for a pair of adsorbed hydrogen atoms to form H$_2$ molecules on grain surfaces. The effective reaction rate $R_H$ for H + H $\to$ H$_2$ is given by \citep{Cazaux_etal_02,Cazaux_etal_04}
\begin{equation}\label{eq:h2}
R_H = \frac{1}{2}s_H n_H v_H n_d \sigma_d \eta\ ,
\end{equation}
where $s_H$ is the sticking coefficient of atomic H on grains (taken to be 1), $v_H$ is the thermal velocity of atomic H, $\sigma_d$ is the collisional cross section between H atoms and dust grains, which is approximately geometric, $n_H$ and $n_d$ are the number densities of atomic hydrogen and dust grains. We adopt $\eta=0.2$ as standard value \citep{Cazaux_etal_04}.

In the disk surface models, we further include far-UV (FUV) radiation, which drives a series of photo-ionization and photo-dissociateion reactions. Accordingly, we extract additional 136 photo-reactions from the UMIST database for disk surface chemistry. We also include photodesorption of CO and H$_2$O. Using Eq. (6) of \citet{Hollenbach_etal_2009}, the flux $F_{pd}$ of adsorbed particles leaving
the surface due to photodesorption is given by

\begin{equation}
    F_{\rm pd} = Y G(r,\Sigma) f_s 
\end{equation}
where $Y=0.001$ for H$_2$O and 0.01 for CO is the yield in molecules per photon. $G(r,\Sigma)$, which is defined in Section \ref{sec:model}, is the photon flux incident on the unit surface area of grain surface and is parameterized in units of the Habing flux $G_0$ \citep{Habing68}. $f_s$ is the surface fraction covered by the molecule. It is however fixed to unity, for simplicity.

We also implement a simplified model to test self-shielding effect for H$_2$ where shielding factor is only a function of surface number density and estimated as $f = \rm min[1, (n/{10^{14} cm^{-2}})^{-3/4}]$ \citep{Wolcott-Green2011}. Shielding for CO and CI cannot be approximated so easily and is not included in this proof-of-concept study. Photoreactions rates for CO and CI are directly adopted from UMIST database. More discussions on the effect of shielding can be found in section \ref{sec:result}. Apart from the grain process listed above, no grain surface chemistry is considered in our models.
 
Some caution should be exercised for ignoring grain-surface chemistry. In the disk surface layers, strong radiation fields are expected to maintain all molecules in the gas-phase and grain surface chemistry is thought to only play a minor role, though explicit tests are needed. In the midplane region, most volatiles except molecular hydrogen exist in the form of icy grain mantles, and hence grain surface chemistry could be important. In particular, grain-surface reactions offer important pathways for forming complex organic molecules (e.g., Walsh et al. 2014). The general outcome depends on a combination of initial chemical conditions, as well as ionization and radiation characteristics.  On the other hand, the disk midplane models we consider start with all O, C and N already in molecular form, making some species less subjective to grain surface chemistry, including water, which mainly form in low temperature (i.e. molecular clouds; \citealp{Cleeves_etal_14}).

\begin{table}
\begin{center}
\caption{``molecular cloud" like initial abundance\label{tab:initial_abundance}}
\begin{tabular}{c|c|c|c}
\tableline
H$_2$ & $5.00 \times 10^{-1}$ & He & $1.40 \times 10^{-1}$ \\
HCN & $2.00\times 10^{-8}$ & CS & $4.00\times 10^{-9}$ \\
CO & $4.33\times 10^{-5}$  & HCO$^+$ & $9.00\times 10^{-9}$ \\
SO & $5.00\times 10^{-9}$ & C$^+$ & $1.00\times 10^{-9}$ \\
N$_2$ & $1.00\times 10^{-6}$ & NH$_3$ & $8.00\times 10^{-8}$ \\
H$_3^+$ & $1.00\times 10^{-8}$ & C$_2$H &$8.00\times 10^{-9}$ \\
H$_2$O(gr) & $2.40\times 10^{-4}$ & CH4(gr) & $5.70\times 10^{-6}$ \\
CH$_3$OH(gr) & $1.80\times 10^{-5}$  & CO$_2$(gr) & $3.30\times 10^{-5}$ \\
N & $2.25\times 10^{-5}$ & CN  & $6.00\times 10^{-8}$ \\
Si(gr) & $1.00\times 10^{-9}$  & Mg(gr) & $1.00\times 10^{-9}$ \\
Fe(gr) & $1.00\times 10^{-9}$ &  & \\
\tableline
\end{tabular}
\end{center}
\end{table}

\subsubsection{Initial Abundances} \label{sec:initial_abundance}

Two types of initial abundances are considered in this paper. The first is ``molecular cloud" like initial abundances, with abundances shown in Table \ref{tab:initial_abundance}. Compared to Table 2 of \citet{Cleeves16}, the only change we made is that we assume metal species are initially adsorbed on grain surface. The second type is the atomic initial abundances, where we convert all species in the ``molecular cloud" initial abundances into their atomic phase. 

For chemistry in the disk midplane, we use the ``molecular cloud" initial abundances since comet observations indicate a large degree of inheritance in disk midplanes \citep{Visser_etal_2011, Taquet_etal_16,Alexander_etal_18}.   At the disk surface, we find that both initial abundances give essentially the same results and we use the atomic initial abundances for simplicity.

\begin{figure}[ht]
\centering
\includegraphics[width =0.45\textwidth]{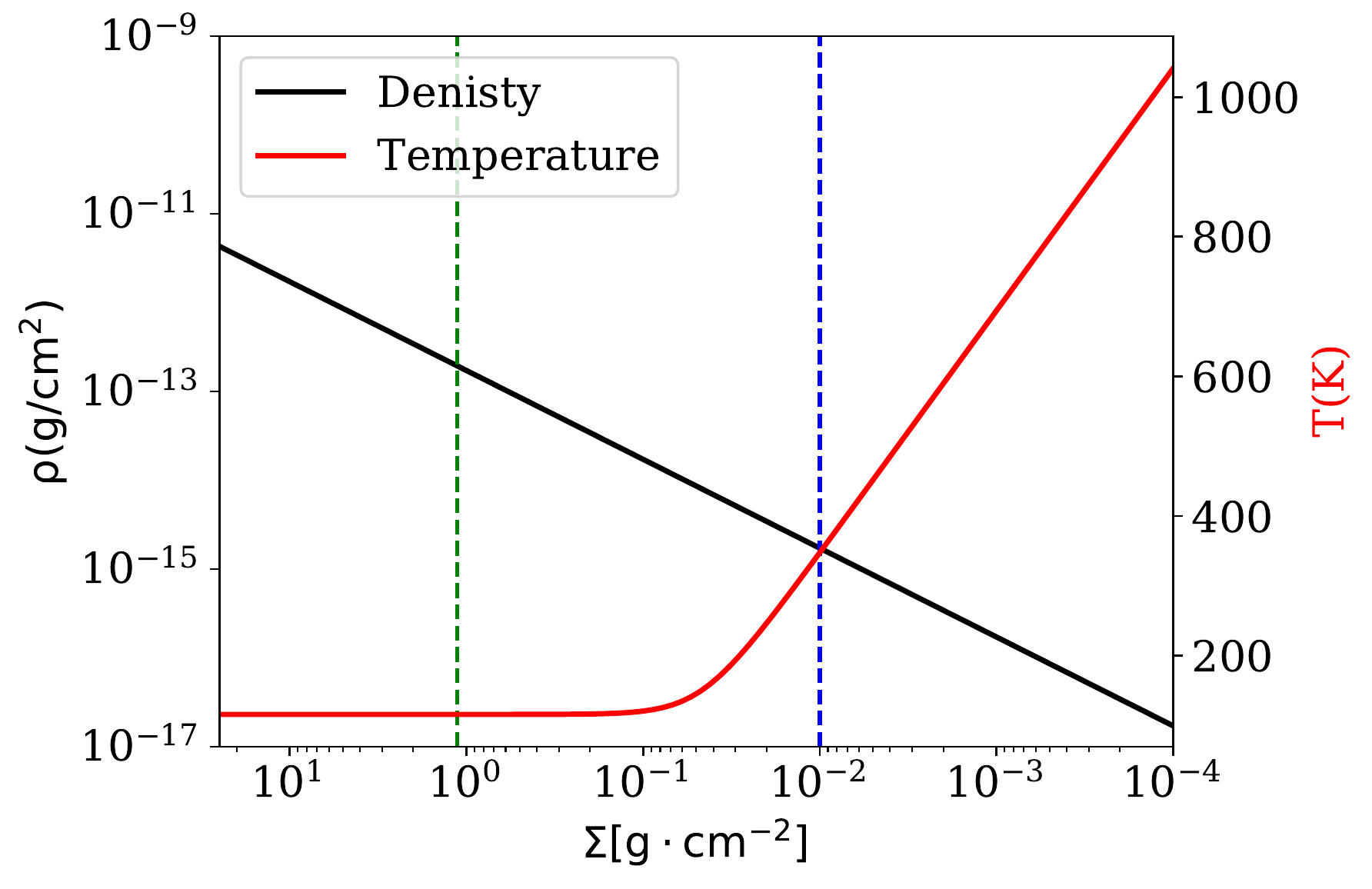}
\includegraphics[width =0.45\textwidth]{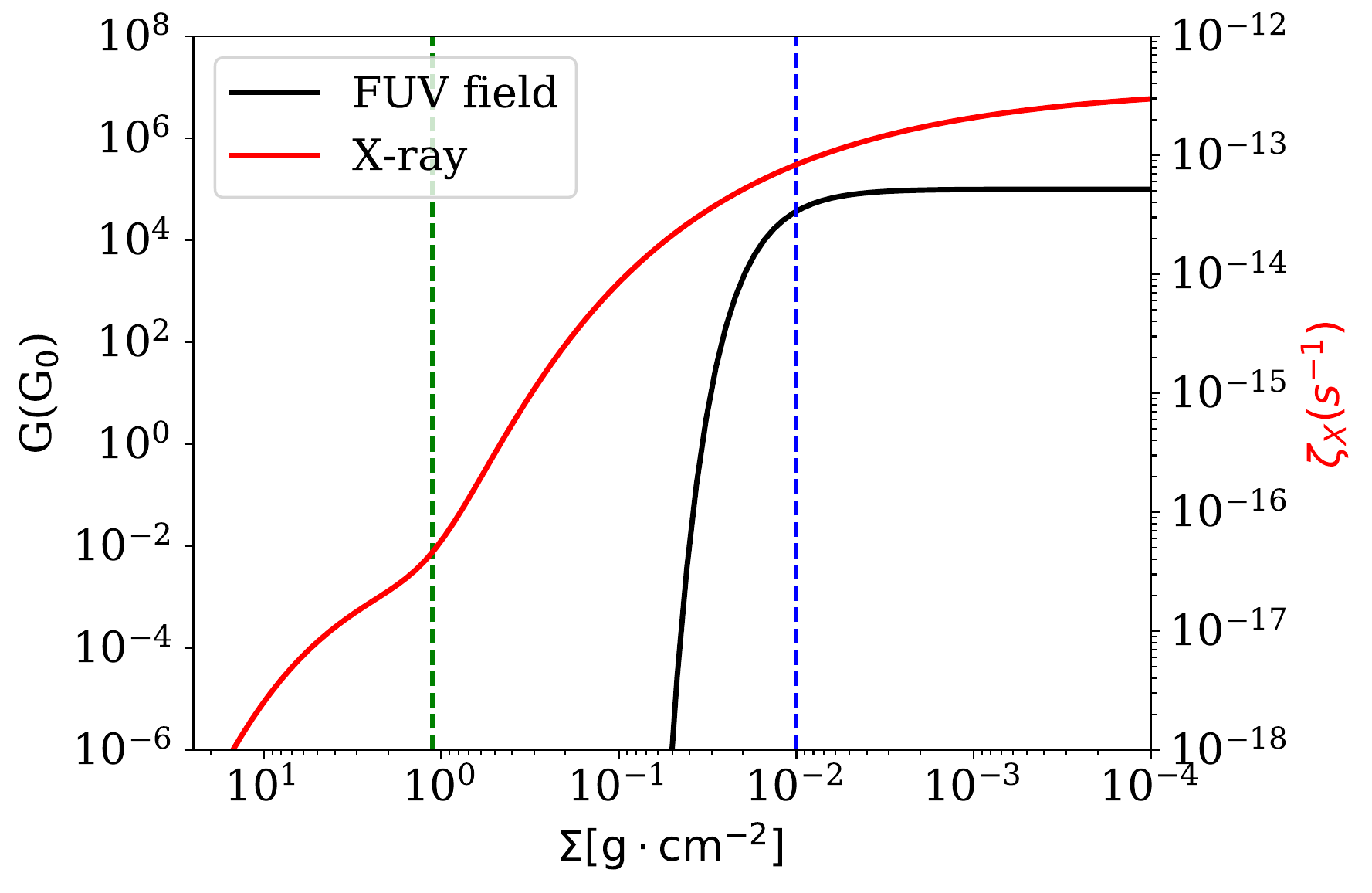}
\caption{{\bf Top:}  Density and gas temperature in the disk surface as a function of surface mass density at 10AU. {\bf Bottom:} FUV radiation field (in the unit of Habing flux) and X-ray ionization rates in the disk surface as a function of surface mass density at 10AU.  Vertical dashed blue line indicates the boundary between surface and intermediate layer in our disk model.  Vertical green line indicates two scale heights above the disk midplane in the disk model (within which the midplane reduced network is valid).} \label{fig:GT}
\end{figure}

\subsection{Disk model}\label{sec:model}

We adopt a thin disk model with surface density profile $\Sigma(r)=500\times r_{\rm{au}}^{-1} \rm{g\cdot cm^{-2}}$ and midplane radial temperature profile $T_{\mathrm{mid}}(r)=200\times r_{\rm{au}}^{-0.6} \rm{K}$ \citep{AndrewsWilliams07,Andrewsetal09}, where $r_{\rm{au}}$ is disk radius measured in AU. Disk is assumed to be vertically isothermal and the vertical density profile is simply determined from hydrostatic equilibrium 
\begin{equation}
\rho(z) = \rho_0 \exp\left(-\frac{z^2}{2H^2}\right)
\end{equation}
where $\rho$ is gas density, $H$ is scale height and defined as $H = c_s/\Omega_K$, where $\Omega_K$ is the midplane Keplerian frequency, $c_s = \sqrt{k_B T_{\rm mid}(r)/\mu m_p}$ is the midplane sound speed, with mean molecular weight taken to be $\mu=2.34$.

Multiple ionization sources are considered in disk midplane. Cosmic-rays give an effective ionization rate of $\zeta_{\rm CR}=10^{-17}$ s$^{-1}$ exponentially attenuated with column density to the disk surface normalized by 96 g cm$^{-2}$ \citep{Umebayashi81}. A spatially uniform ionization rate by radioactive decay is set to $7.0\times 10^{-19}$ s$^{-1}$. For X-ray ionization $\zeta_X$, a fitting formula from \citet{Bai09} to \citet{Igea99} is used by assuming X-ray temperature $T_X$ = 3 KeV and X-ray luminosity $L_X$ = 10$^{30}$ erg s$^{-1}$. 

\begin{equation}
\begin{split}
\zeta_X &=\frac{L_X}{10^{29}{\rm erg/s}}r_{\rm au}^{-2.2} [\zeta_1\left( e^{\rm -(N_{H1}/N_1)^\alpha} + e^{\rm -(N_{H2}/N_1)^\alpha} \right) \\
& + \zeta_2\left( e^{\rm -(N_{H1}/N_2)^\beta} + e^{\rm -(N_{H2}/N_2)^\beta} \right) ] \\
\end{split}
\end{equation}

where N$_{\rm H1,2}$ is the column density of
hydrogen nucleus vertically above and below the point of interest, $\zeta_1 = 6.0\times 10^{-12} s^{-1}$, N$_1$ = $1.5\times 10^{21} {\rm cm^{-2}}$, $\alpha$ = 0.4, $\zeta_2 = 1.0\times 10^{-15} s^{-1}$, N$_2$ = $7.0\times 10^{23} {\rm cm^{-2}}$, and $\beta = 0.65$.

\subsubsection{Disk model in the surface layer} \label{sec:disk_surface}

 In this paper, the surface layer is the defined as the region where FUV penetration is high (attenuation is less than $e^{-1}$). This layer is characterized by a photon-dominated chemistry and strong gas-heating. We prescribe a simple gas temperature profile that increases logarithmically with decreasing surface density $\Sigma$ (i.e.,  column density vertically above the point of interest) as
\begin{equation}
T_g (r,\Sigma) = T_{\mathrm{mid}}(r)\log \left[10+\left(\frac{0.1\mathrm{g\cdot cm^{-2}}}{\Sigma}\right)^3 \right]\ .
\end{equation}
Density at disk surface always deviates from Gaussian due to the rapidly varying vertical temperature profile, magnetic pressure support and disk wind. For our purposes, it suffices to simply approximate the gas density by $\rho = \Sigma/H$\footnote{At the surface, mean molecular weight varies as more molecules are destroyed, but for our purpose it suffices to approximate it as a constant.}. With photochemical reactions being the key ingredients in disk surface chemistry, these approximations should suffice for the purpose of network reduction.

FUV radiation is characterized by a small penetration depth of $\sim0.01-0.1$ g cm$^{-2}$ followed by rapid attenuation \citep{Perez-Becker11}, and hence is neglected in disk midplane network. Attenuation of FUV results mostly from very small dust grains (e.g. PAHs), whose properties, especially abundances, are highly uncertain \citep{Geers_etal_2006,Geers_etal_2009,Keller_etal_2008}. We parameterize the FUV fluxes at the disk surface as a function of radius and column density to disk surface $\Sigma$ to mimic the presence of tiny dust grains
\begin{equation}
G(r,\Sigma)= \frac{10^7}{r_{\rm au}^2} \exp\left[-\left(\frac{\Sigma}{\Sigma_{\rm FUV}}\right)^2\right]\ ,
\end{equation}
where the value $10^7$ at 1 AU is considered typical for T Tauri stars \citet{Bergin_etal_2004}, and we adopt $\Sigma_{\rm FUV}=0.01\mathrm{g\cdot cm^2}$ as the characteristic FUV penetration depth (assumed to be a constant). While not rigorous, this prescription simply aims to mimic the attenuation of FUV radiation over height from more realistic calculations 
(e.g., \citealp{Nomura05,Walsh_etal10}), and $\Sigma_{\rm FUV}$ could be larger for lower tiny grain/PAH abundances. The general properties at a representative radius of $R=10$ AU in the surface region are summarized in Figure \ref{fig:GT}.

\begin{figure}[ht]
\centering
\includegraphics[width =0.45\textwidth]{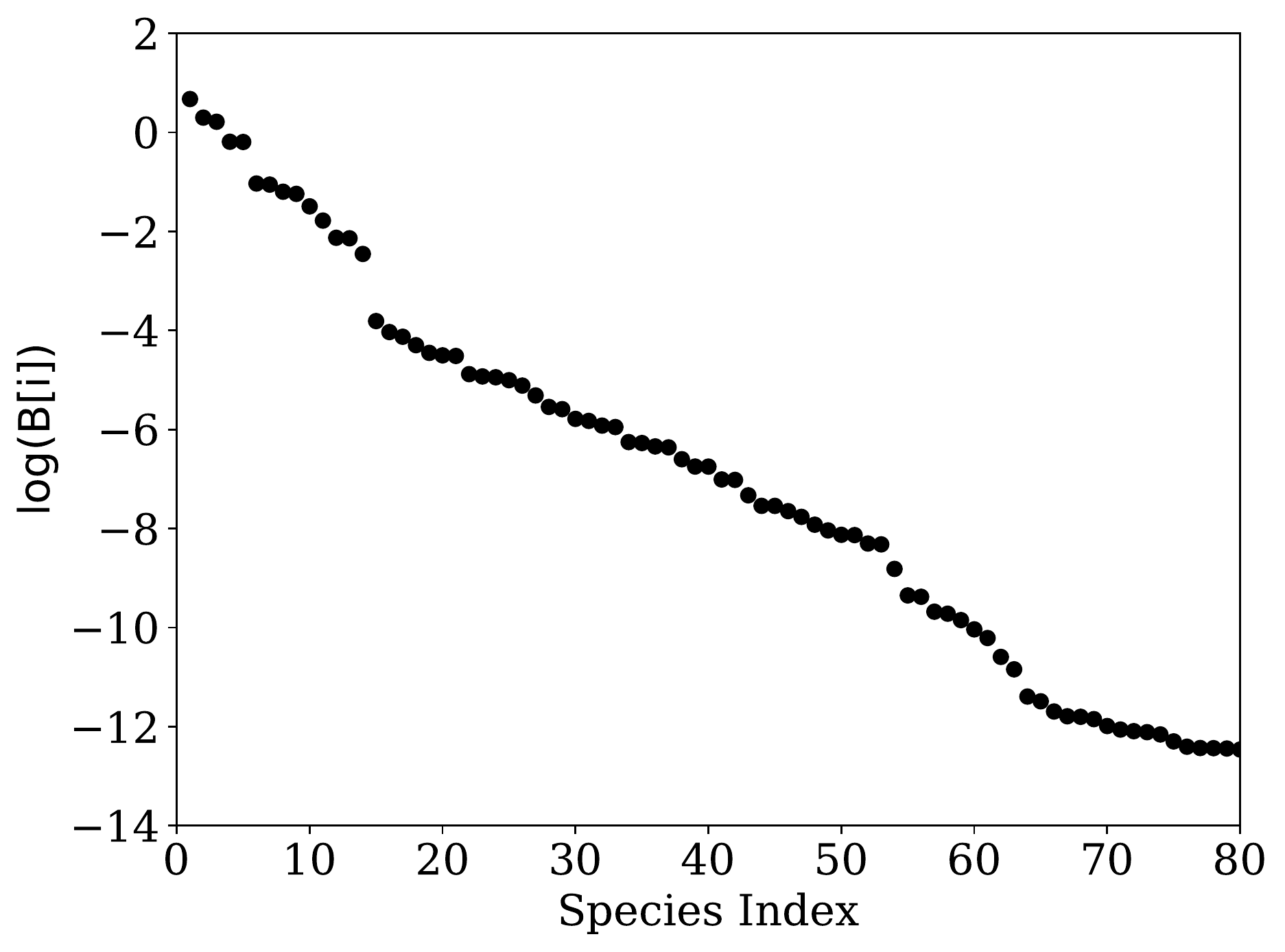}
\includegraphics[width =0.45\textwidth]{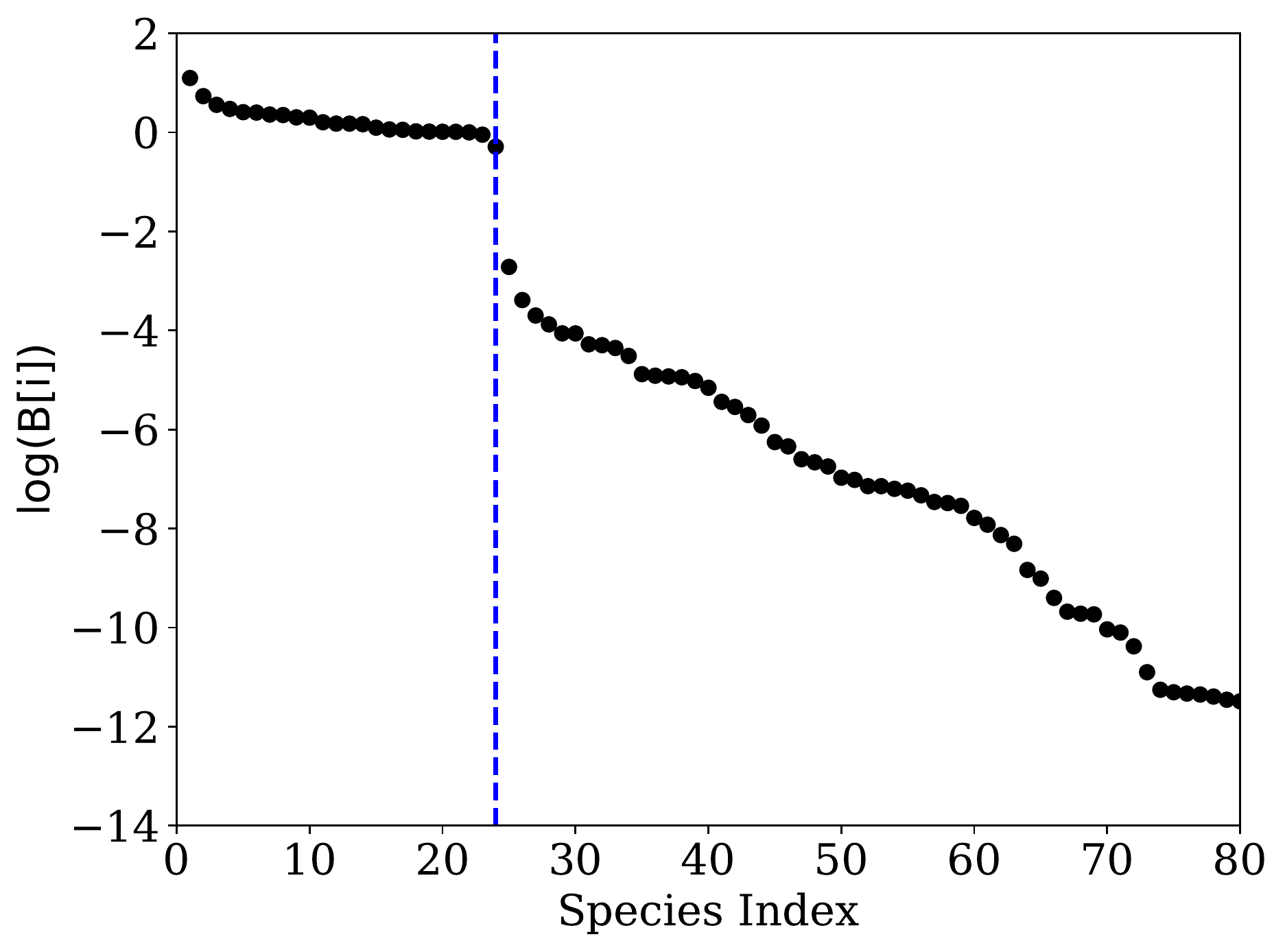}
\caption{Sorted sensitivity B[$i$] for all species in the full network in disk midplane at 1AU, after 5 iterations (top panel) and after the identification of necessary species is completed (22 iterations, bottom panel). The dashed line in the bottom panel indicates the division between species in the reduced network (necessary species) and outside the reduced network (unnecessary species).  Also see Figure 1 in \citet{Wiebe03}.}\label{fig:weight}
\end{figure}

\subsection{Reduction method}\label{sec:reduction}

We evolve the full network for a certain time (i.e., 1 Myr), and the reduction is performed based on the species abundances and reaction rates at this time. While in principle the reduction should be performed at different times to capture different stages of chemical evolution, we show in Appendix \ref{appen:A} that the reduced network constructed at the last time step agrees well with full network throughout almost the entire evolutionary process. This time dependence can be, to some extent (though not self-consistent), considered as to mimic the effect of advection when coupled with dynamics \citep{semenov04}, where fluid advection keeps mixing new species into the system to bring it out of equilibrium.

Our network reduction method is iterative and is described as follows.
First, we specify a number of species whose abundances we are most interested in. They serve as initial species of the reduced network and new species will be added iteratively following our procedures. The criteria for adding new species is controlled by the sensitivity parameter B[$i$], which measures the importance of the $i$th species to species already selected in the reduced network. The definition of $B[i]$ reads \citep{Tomlin92,Ruffle02}
\begin{equation}
B[i] =\sum^{N^\prime}_{j=1}\left(\frac{n_i}{g_j}\frac{\partial f_j}{\partial n_i}\right)^2
\end{equation}
where the summation over $j$ is restricted to all species selected in the current iteration of the reduced network (containing $N$' species), $n_i$ is the number density of the $i$th species, $f_j$ is the {\it net} formation or destruction rate of the $j$th species, and $g_j$ is the larger one between {\it total} formation and {\it total} destruction rates of the $j$th species (without adsorption and desorption). The term inside the bracket represents the sensitivity of the net production/destruction rate of the $j$th species to the abundance of $i$th species. After the summation, $B[i]$ indicates the sensitivity of the reduced network to the abundance of $i$th species. Larger $B[i]$ means that the $i$th species is more important for determining the abundances of the species in the reduced network. Sensitivity $B[i]$ is calculated only for gas phase species and dust grains but not for mantle species, because mantle species desorption rate can be very temperature sensitive and may overwhelm the $B[i]$ parameter and adversely affects the reduction process. On the other hand, mantle species will automatically be included if their gas-phase counterparts are selected.

After computing $B[i]$ for each species, we sort all species according to their $B[i]$. If the ratio between the smallest value of $B[i]$ within the reduced network and the largest value of $B[i]$ outside the reduced network is higher than some threshold (see below), we consider that the currently selected set of species is sufficiently self-contained, and the identification of necessary species is complete. Otherwise, we incorporate one new species with the largest $B[i]$ into the network (i.e., the most necessary new species), and proceed to the next iteration.

\begin{table}
\begin{center}
\begin{threeparttable}
\caption{Reduced network in disk midplane\label{tab:mid}}
\begin{tabular}{c|l}
\tableline   
Neutrals &  H H$_2$ H$_2$O He  N$_2$  CO CH$_4$ CO$_2$  O$_2$ NH$_3$ \\
\tableline
\multirow{2}{*}{Ions}&  e$^-$ H H$_2^+$ H$_2$O$^+$ He$^+$  N$_2^+$  CO$^+$  CH$_2^+$ CH$_3^+$ CH$_4^+$  \\
    & CH$_5^+$  H$_3^+$  H$_3$O$^+$ NH$_4^+$ C$^+$ OH$^+$ HCO$^+$ N$_2$H$^+$    \\
\tableline
\end{tabular}
\end{threeparttable}
\end{center}
\end{table}

To illustrate this procedure, we show in Figure \ref{fig:weight} the sorted distribution of sensitivity B[$i$] in disk midplane at 1AU after 5 iterations (top panel) and after 22 steps when the iteration is completed (bottom panel, see \S  \ref{sec:midplane} for more detail). As more species are included, the difference in $B[i]$ between the selected species and the remaining species becomes more and more prominent, and eventually, a clear cutoff is developed. Quantitatively, we find that 2 to 3 orders of magnitude drop in sensitivity between the least necessary species within the reduced network and the most necessary species outside the reduced network is a good criterion for terminating the iteration.

\subsubsection[]{Application to Our Disk Model}

We run network reduction at 1, 10 and 100 AU at both the midplane, and the disk surface with $\Sigma=\Sigma_{\rm FUV}=0.01$ g cm$^{-2}$. We expect these locations to cover broad disk environments in the midplane region (within $\sim2H$) as well as the surface layer ($\Sigma\lesssim\Sigma_{\rm FUV}$). Note that the intermediate layer in between is neglected in this work. The resulting six reduced networks will then be combined into two. One for the midplane and one for the surface.

The initial reduced network we choose in both disk midplane and disk surface contain electrons, because ionization fraction is an important factor for gas dynamics. In the midplane region, we also incorporate CO in the initial reduced network. Besides being an important gas tracer, CO is also directly related to HCO$^+$, which is a dominant charge carrier and is crucial for disk ionization balance. In the disk surface, we also incorporate major coolants (i.e. CO, H$_2$O, Si, S), whose abundances determines the heating/cooling rates that crucially affect disk mass loss and observational diagnostics. The chemistry of Si and S in the disk surface is dominated by photoreactions and is mostly isolated from other species. Removing them from initial reduced network would in general only affect species directly associated with them, e.g. their ionized counterparts.

\begin{figure}[ht]
\centering
\includegraphics[width =0.45\textwidth]{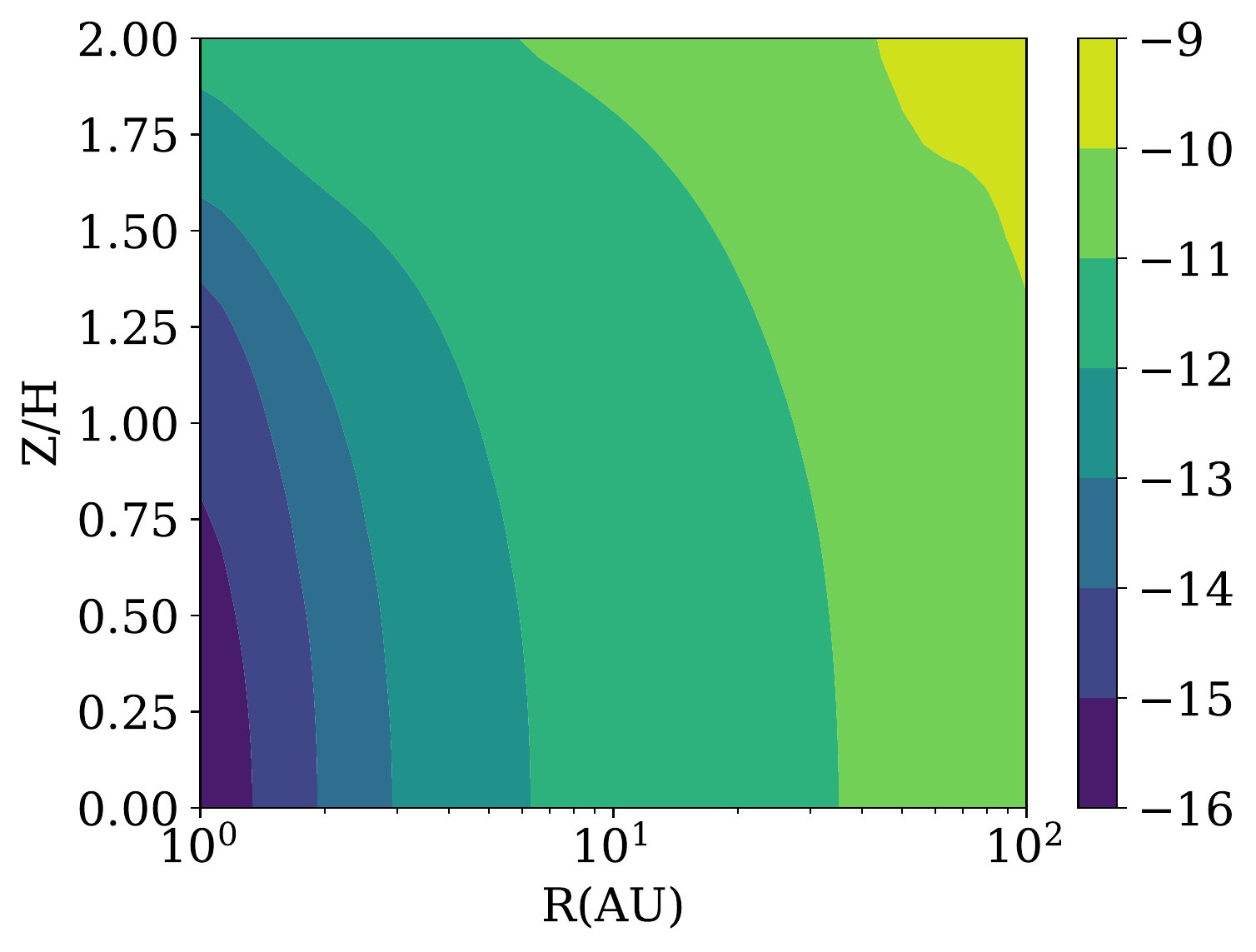}
\includegraphics[width =0.45\textwidth]{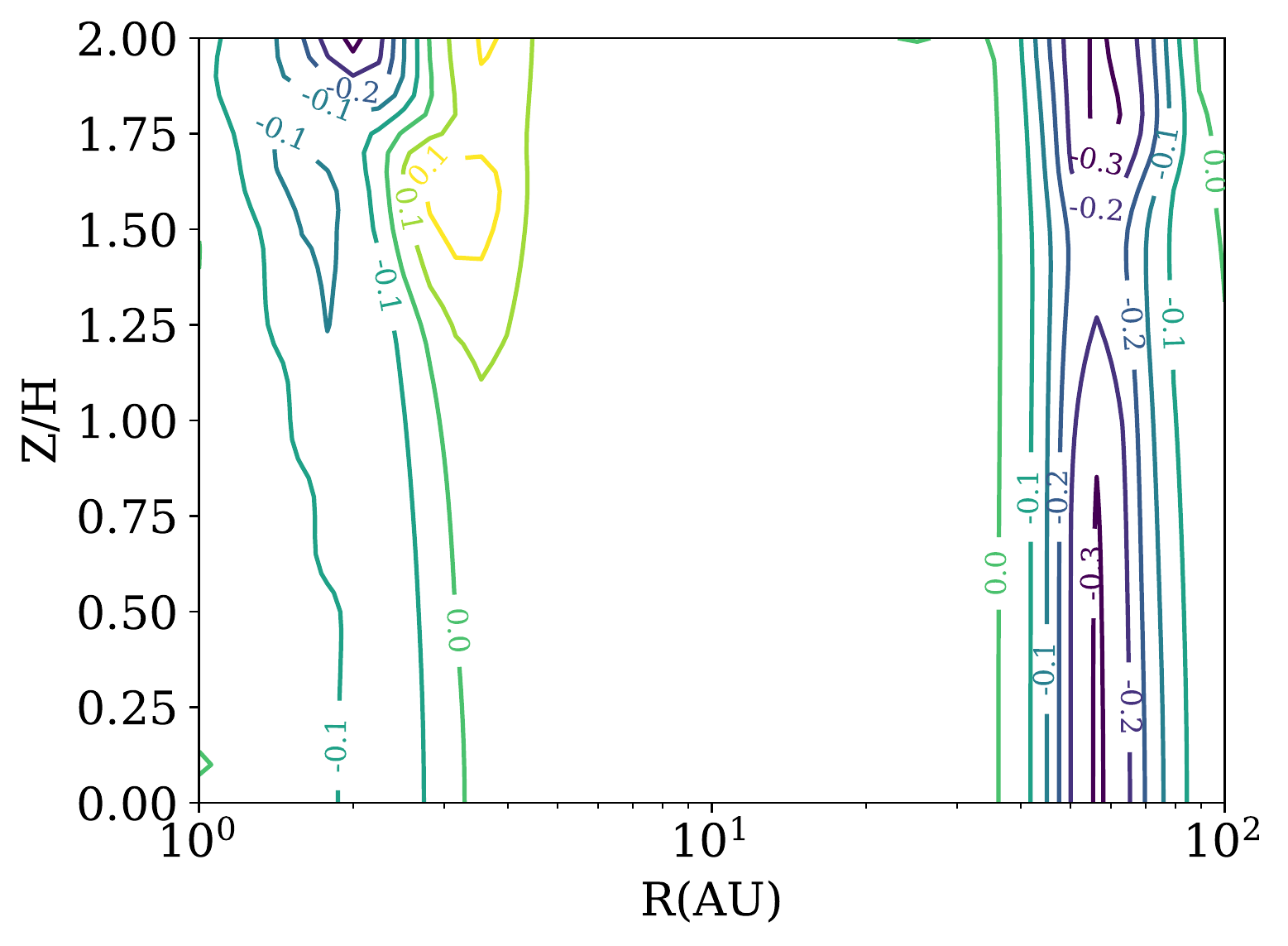}
\caption{{\bf Top}: Ionization fraction (relative to total H atoms) in logarithmic scale as a function of radius and vertical height from the full network after 1Myr evolution. {\bf Bottom}: $\log\left(n_{\rm e, reduced}/n_{\rm e,full}\right)$ as a function of spatial location after 1 Myr evolution, where $n_{\rm e, reduced}$ are $n_{\rm e, full}$ are the electron number densities from the reduced and full network, respectively.}
\label{fig:2denum}
\end{figure}

Once the species are selected at different radii and combined together, a reduced network can be extracted from full network by keeping all reactions that only involve these species. The next step is a sensitivity analysis based on reaction rates to eliminate redundant reactions (see \citet{Tomlin92} for more detail). Here we only keep important formation and destruction pathways based on their reaction rates. For each species in the combined reduced network, we only choose the dominant formation and destruction reaction channels with their total reaction rates exceeding 99\% percentage of total formation/destruction rates. Dominant reactions are selected at different radii and then combined together for disk midplane and surface regions, respectively.

\section{Results} \label{sec:result}

We apply the technique described in previous section and construct reduced chemical networks. In \S \ref{sec:midplane}, we discuss network reduction in the disk midplane regions, and in \S \ref{sec:surface}, we apply the same technique at disk surface dominated by photochemical reactions. In \S \ref{sec:compare}, we compare the results with previous works.

\subsection{Network reduction in disk midplane}\label{sec:midplane}

Electron and CO are used as our initial reduced network. In disk midplane, 22, 23, and 10 gas-phase species are selected at 1AU, 10AU, 100AU separately. Adding these necessary species together and keeping only dominant reactions involved by these species, the final reduced network has 28 species and 53 gas-phase reactions (see Table \ref{tab:mid} for the species list and Table \ref{tab:mid_reaction} in Appendix \ref{appen:reactionlist} for the reaction list). 

\begin{figure}[ht]
\centering
\includegraphics[width =0.45\textwidth]{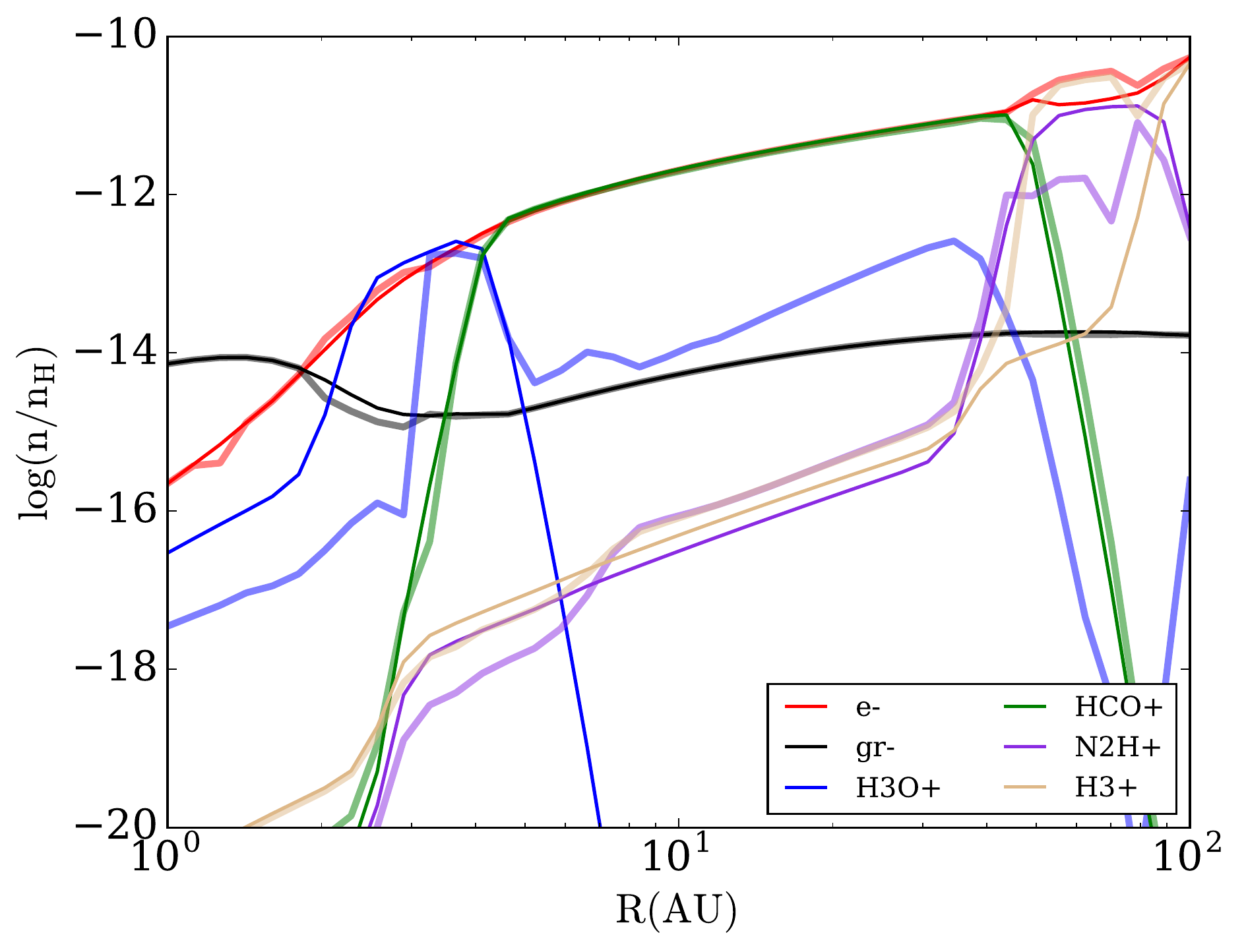}
\includegraphics[width =0.45\textwidth]{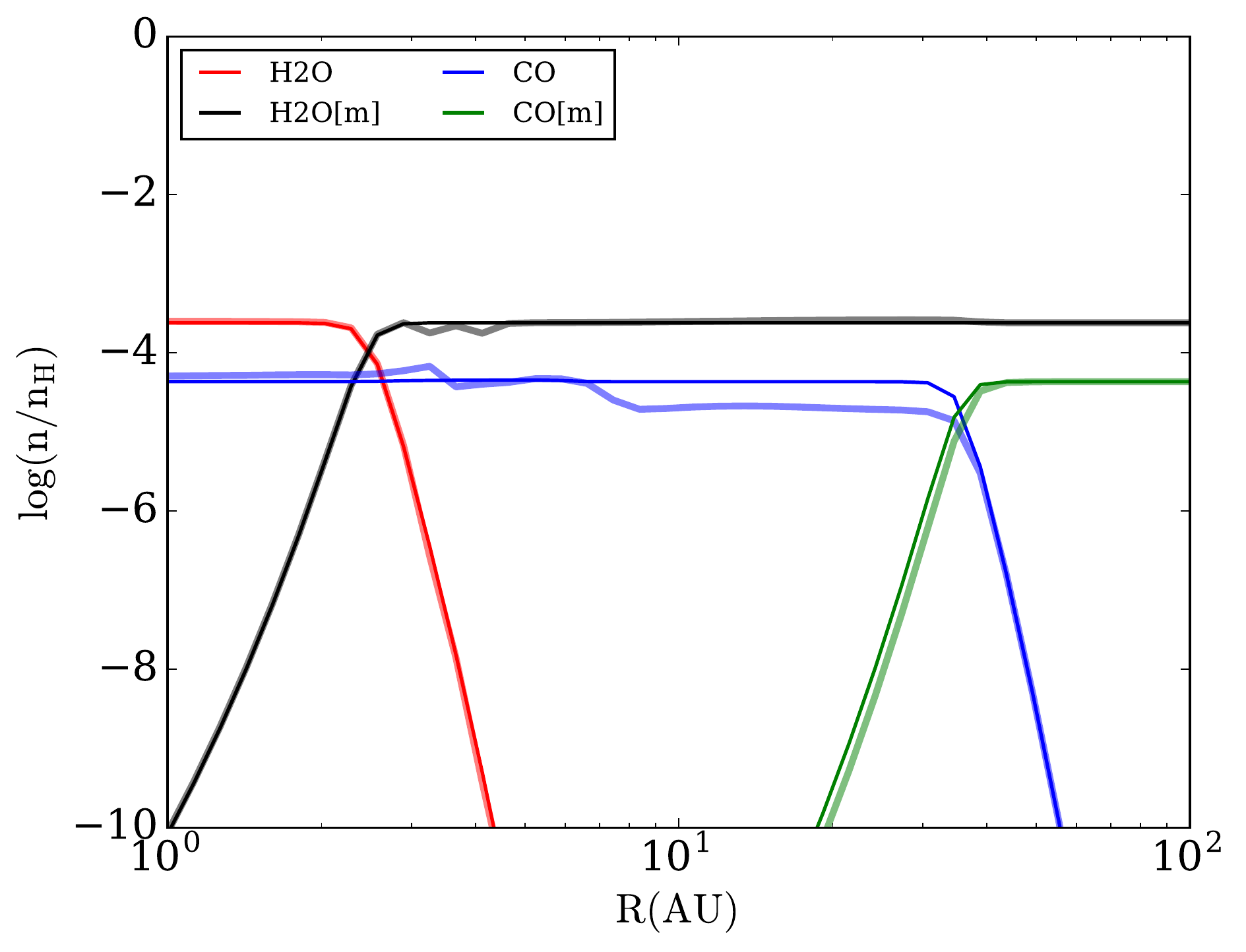}
\caption{Abundances of electrons and dominant ions (top panel) and neutral species (bottom panel) in the reduced network (thin lines) compared to the full network (thick lines) at the disk midplane. [m] indicates mantle species of their gas phase counterparts.}
\label{fig:numden}
\end{figure}

Primary carriers for charges, hydrogen, oxygen, carbon and nitrogen are all selected in the reduced network. In Figure \ref{fig:2denum}, we show the disk ionization fraction from the full network (top panel) and log(n$_{\rm e,reduced}$/n$_{\rm e,full}$) (bottom panel) as a function of spatial locations after 1 Myr' evolution, where $n_{\rm e, reduced}$ are $n_{\rm e, full}$ are the electron number densities from the reduced and full network, respectively.  The ionization fraction from our chemical network is consistent with previous works by \citet{Ilgner06,Bai09} (e.g. See Figure 4 in \citealp{Bai09}). The ionization fraction in the reduced network agrees with full network very well within 2 scale heights above/below the midplane. Differences are well within a factor of 2, and are within $\sim10\%$ in most regions. The largest deviation occurs at higher altitude and within a radius of a few and tens of au. This is not surprising given that original network reduction was conducted at the disk midplane. While the deviations can be reduced by incorporating more species, we also note that the chemistry in these upper layers (i.e., the intermediate layer) is intrinsically more complex (SWH04), which we leave for future work for more detailed investigation that include grain surface chemistry in the original network, and potentially a different reduction scheme.

\begin{figure*}[ht]
\centering
\includegraphics[width =0.9\textwidth]{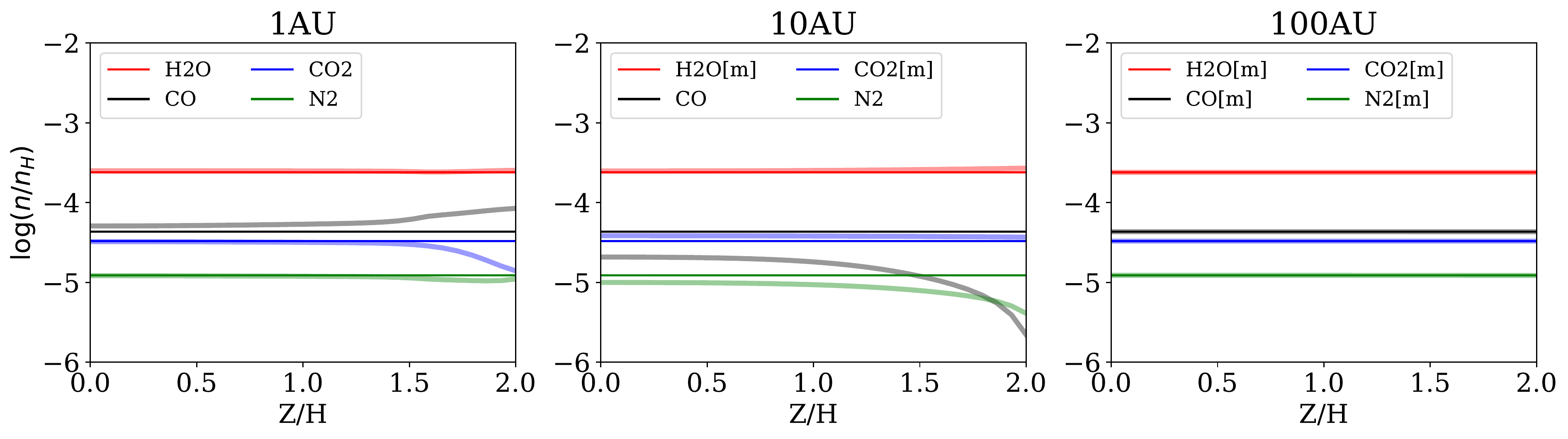}
\includegraphics[width =0.9\textwidth]{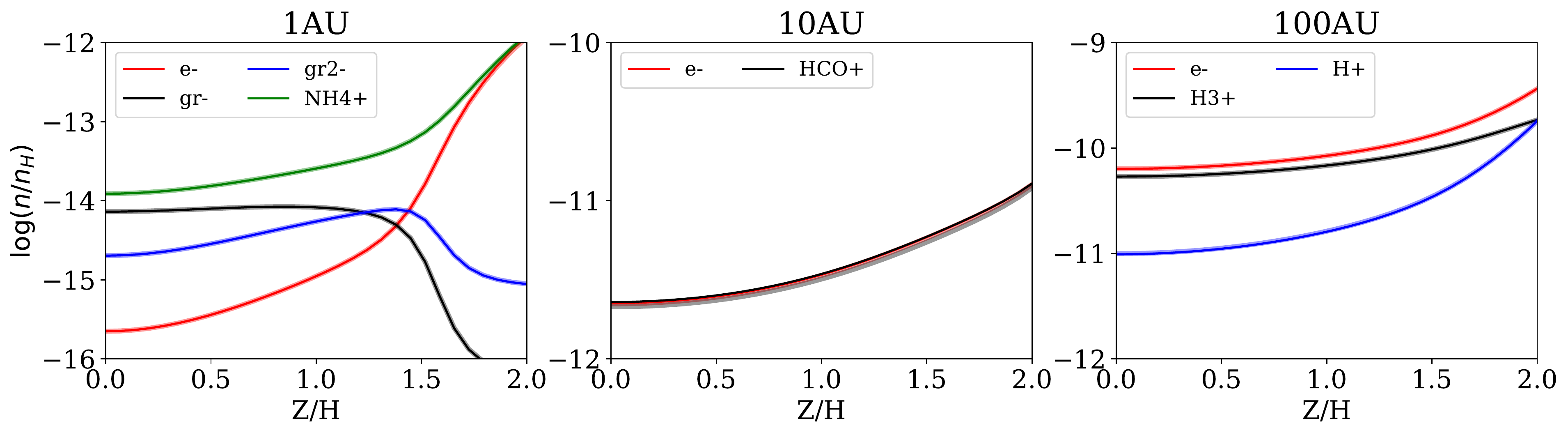}
\caption{Vertical profile of dominant neutrals (top panels) and ions (bottom panels) number density (relative to total H atoms )  in reduced network (thin lines) and full network (thick lines) at 1AU (left panels), 10AU (middle panels) and 100AU (right panels).  [m] indicates mantle species of their gas phase counterparts.}
\label{fig:vden}
\end{figure*}

Besides electrons, the final reduced network also yields relatively accurate abundances for most other species in the reduced network. Figure \ref{fig:numden} shows the comparison between reduced network (thin lines) and full network (thick lines) for several ions (top panel) and neutral species (bottom panel) as a function of radius. In the top panel, calculation from the full network shows that dust grains are dominant charge carriers in inner disk ($\sim$ 1AU). This is replaced by H$_3$O$^+$ at $\sim$2AU.  In between a few AU and $\sim40$ AU (where the CO snow line is located), HCO$^+$ is the dominant ion. Beyond the CO snowline, N$_2$H$^+$ abundance increases rapidly and takes over to become the dominant ion, see also \citep{Qi_etal13}. Near 100AU, H$_3^+$ replaces N$_2$H$^+$ and becomes the dominant charged particle. Metal elements Fe and Mg are not present in the reduced network because they almost entirely freeze out onto the grains in the temperature range explored. In a grain-free network where metals are dominant charge carriers, we tested that they will be selected in the reduced network, as expected.

 Accurately capturing the ice lines of volatile species is important in many aspects of planet formation (e.g., \citealp{Oberg2011,Madhusudhan11}), whose exact locations are varing throughout PPDs lifetime (e.g. \citealp{Piso_etal_15}). In the bottom panel of Figure \ref{fig:numden}, we see a clear transition from gas phase water to water ice at $\sim$2-3AU, gas phase CO to CO ice at $\sim$40AU both in full network and reduced network. This is not surprising because the transition is primarily determined by the balance between desorption and sublimation.  

\begin{table}
\begin{center}
\caption{Reduced network in the disk surface\label{tab:surface}}
\begin{tabular}{c|l}
\tableline
Neutrals &  H C O OH H$_2$O CO  Si S  H$_2$ \\
\tableline 
\multirow{2}{*}{Ions}& e$^-$ H$^+$ H$_2^+$ C$^+$ O$^+$ OH$^+$ CO$^+$ H$_2$O$^+$ Si$^+$ S$^+$ \\
& CH$^+$  H$_3$O$^+$ SiO$^+$ HCO$^+$ \\        
\tableline
\end{tabular}
\end{center}
\end{table}

We also compare the vertical profiles of the dominant neutral and ion species in the reduced network (black lines) with those in the full network (red lines), as shown in Figure \ref{fig:vden}. In the top three panels, we see that dominant neutrals show very good agreement with full network at 1 au and 100 au. At 10 au, and above one scale height at 1 au we do see differences on the order of a few for several species, and at 10 au the discrepancy is above an order of magnitude for CO at the most elevated disk heights considered. The dominant ions, as shown in bottom three panels, are accurately reproduced by the reduced network at all radii and vertical altitudes.

\begin{figure*}[!ht]
\centering
\includegraphics[width =1.0\textwidth]{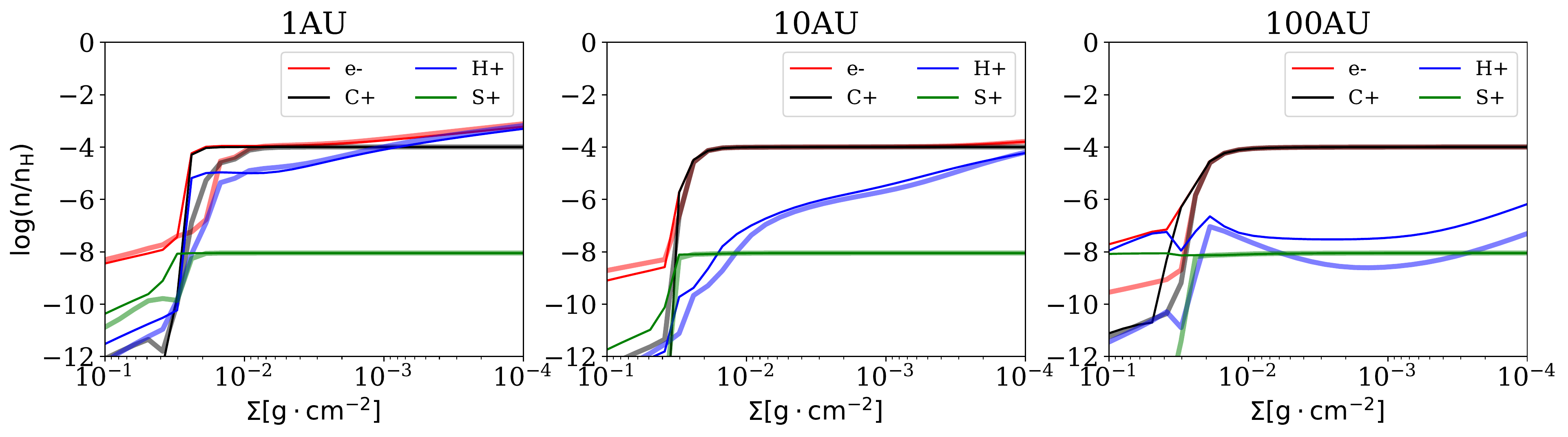}
\includegraphics[width =1.0\textwidth]{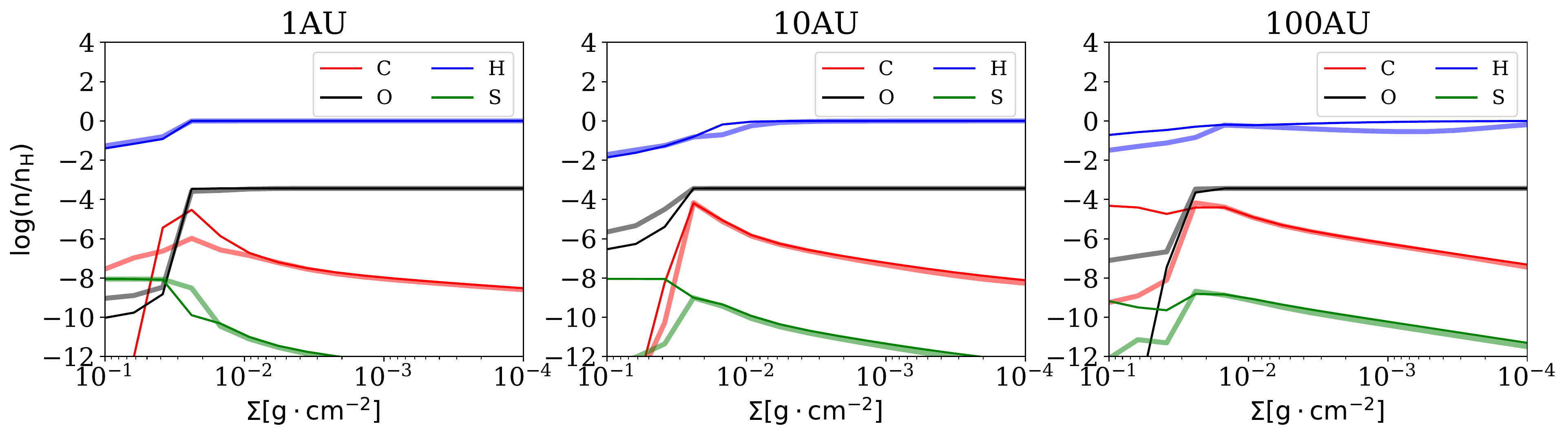}
\includegraphics[width =1.0\textwidth]{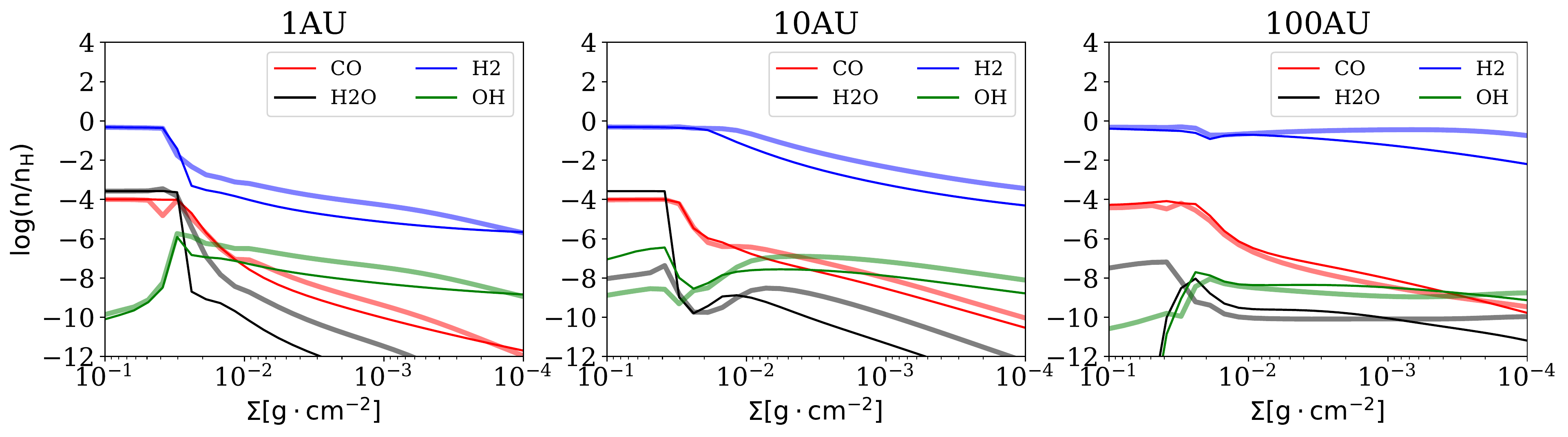}
\caption{Logarithmic scale of ion number densities (top panels), neutral atoms number densities (middle panels) and molecule number densities (bottom panels) as a function of surface mass density (the column density of hydrogen nucleus vertically above the point of interest) from the full network (thick lines) and the reduced network (thin lines) at 1AU (left panels), 10AU (middle panels) and 100AU (right panels).}
\label{fig:sden}
\end{figure*}

\subsection{Network reduction at disk surface}\label{sec:surface}

In the disk surface, with prescribed vertical temperature profile, FUV radiation field and additional photoreactions, we perform network reduction at the FUV front where surface mass density is given by $\Sigma=\Sigma_{\rm FUV} = 0.01 {\rm g\cdot cm^{-2}}$. Important coolants i.e. CO, Si, S, and H$_2$O are adopted together with electron as initial species in the reduced network (e.g. \citet{WangGoodman17}). By performing the reduction, 13, 16, 16 gas-phase species are selected at 1 AU, 10 AU and 100 AU separately. Merging the results, the final reduced network contains 23 gas phase species and 58 gas phase reactions including photoreactions (see Table \ref{tab:surface} for the species list).

Top three panels in Figure \ref{fig:sden} show fractional ionization and dominant ions as a function of surface mass density $\Sigma$ at different radii. We see that the ionization fraction first increases dramatically with decreasing $\Sigma$ and approaches a constant value $\sim 10^{-4}$ when $\Sigma$ is around a few times $10^{-2}\rm{g\cdot cm^2}$, as carbon becomes fully ionized. For even lower $\Sigma$ and higher temperature, ionization fraction tends to increase further as H$^+$ replaces C$^+$ to become the dominant ions.  At all radii, the reduced network  well reproduces the ionization fraction and species abundances when $\Sigma\lesssim\Sigma_{\rm FUV}$, i.e. abundances are within a factor of a few with a handful of exceptions.  

At higher densities where UV radiation is shielded, the reduced network does not accurately reproduce the ionization fraction because more reactions involving polyatomic species become progressively more important. The chemistry in this intermediate layer is more complex and a separate network reduction is needed. We also conducted a separate network reduction  at $\Sigma=0.1$g$\cdot$cm$^2$ where the FUV field is largely shielded. We find that $\sim$ 80 species are required to accurately capture the bulk chemistry in this region, even if only gas-phase chemistry is considered. Similar conclusion was also reached by SWH04.  Besides, gas-grain chemistry and non-thermal desorption beyond photodesorption, which we have ignored, are expected to play an increasingly important role in this layer.

Self and mutual-shielding play an important role in the disk surface. We tested the reduction from both within and without H2 self-shielding implemented and obtain the same network. While CO and CI self-shielding are not implemented, we tested reduction with CO photodissociation and CI photoionization turned off (the limiting case assume very effective self-shielding) and again obtained roughly the same network, with only a few different species\footnote{More rigorously, which is not proceeded here, one could run network reduction with simplified prescriptions of self and mutual-shielding (e.g., using varied $A_v$ values), and then compare the results between full and reduced networks by implementing more complex shielding functions of, e.g., Visser et al. (2009).}. These results suggest that the reduced network is reasonably robust and could also be used in models with more accurate self-shielding prescriptions.

Our reduced network well reproduce the abundances of major coolants such as O, OH, CO, Si, and S in the (UV) photon-dominated surface layer, as shown in the middle and bottom panels of Figure \ref{fig:sden}. Among these major species, only H$_2$O abundance shows relatively large deviation from the full network results by one order of magnitude at $\Sigma \le 10^{-2} \rm g\cdot cm^{-2}$. This occurs when H$_2$O is no longer the dominant species in these regions, which then carries less weight based on the reduction method. This layer largely covers the bulk and base of the (magneto-)photoevaporative flow \citep{Perez-Becker11,Bai_etal16,Wang_etal_17,Nakatani_etal_17}. Therefore, we expect our reduced network to provide accurate results for simulating such outflows when coupled with dynamics, and it will also help provide important information for observational diagnostics.

\subsection{Necessity of Time-dependent Chemistry}\label{sec:timescale}

The importance of treating the chemistry time-dependently rather than assuming steady state can be assessed by comparing chemical and dynamical timescales: time-dependent chemistry is necessary in dynamical simulations when relevant chemistry timescales exceed the dynamical timescale.  

The chemical timescale can vary dramatically from species to species and from reactions to reactions. Here, for our purposes, we give a simple definition that may be considered as a rough estimate for the bulk chemical evolution as follows. We first evolve the full chemical network for 10 million years. We then define the timescale as the time when half of the species’ abundances reaches within 10\% compared to their final abundances. Also, we can separate out the ionization timescale, which can be similarly defined as the time the ionization fraction reaches to within 10\% of the final ionization fraction.

\begin{figure}[!ht]
\centering
\includegraphics[width =0.45\textwidth]{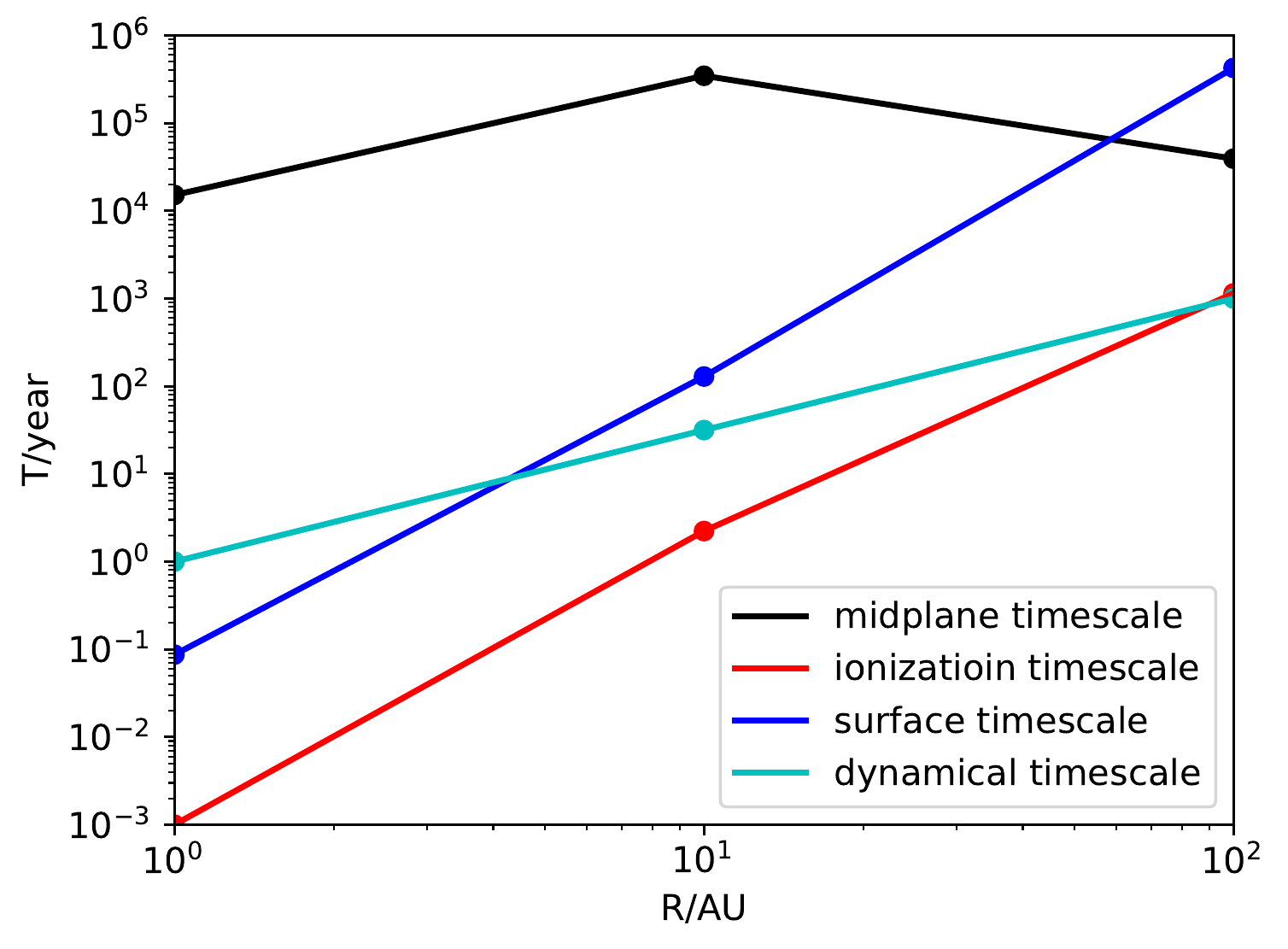}
\caption{Chemistry timescale in the disk midplane (Z=0, black line) and in the disk surface (where we take $\Sigma$ = $10^{-2}$g$\cdot$cm$^{-2}$, blue  line) compared to ionization timescale in the disk midplane (red line) and disk orbital timescale (cyan line).}
\label{fig:timescale}
\end{figure}

In Figure \ref{fig:timescale}, we show the bulk chemistry timescale at different radii at disk midplane (black line) and disk surface (blue line), together with the ionization timescale in the disk midplane (red line). Orbital (dynamical) timescale (cyan line) is also plotted. Chemistry timescale in the disk midplane is dominated by neutral gas reactions, while disk surface chemistry timescale is dominated by photochemistry. Although photochemistry is a rather rapid process, volatile species could be gradually converted to less volatile species and frozen onto grains, which results in longer chemical timescales, especially in the outer disk region\footnote{The same calculation for the grain-free case shows that the inner disk chemistry timescale is mostly unaffected while the chemistry timescale at 100AU is reduced by ~2 order of magnitude, which is still larger than dynamical timescales.}. We see that for both regions, the timescale is either comparable or larger compared to orbital timescales, and hence justifying the importance of coupling gas dynamics with chemistry (the only exception being the the surface at 1 au).  Compared to neutral species, ionization fraction reaches steady state in relatively smaller timescales, especially in the inner disk. However, in the outer disk (e.g. 100AU), ionization timescale is also comparable with local orbital timescale.

\subsection{Comparison with previous work}\label{sec:compare}

 Our results are generally consistent with the work done previously by SWH04, who started from UMIST95 database, when comparing our results to theirs at the specific radii and disk heights they modeled. The size of our reduced network is similar with SWH04 if we combine their reduced networks at different radii together. They focused on reproducing the chemistry at individual disk locations through location-specific reductions. However, it is not clear how well such a network would function at the regions in between where the individual reduced networks were derived. Our reduced network was developed to work for either the entire midplane or surface, and we have verified that ours agree with the full network for the entire midplane or surface.

In the disk surface, our reduced network is larger than SWH04 for their X-ray and UV dominated chemistry (see Table 7 and 8 in SWH04). This is primarily because besides ionization fraction, we also want to keep dominant coolants in the reduced network and accurately capture their abundances. In the intermediate region where FUV radiation is shielded, SWH04 also claimed that the chemistry is the most complex and in some cases more than 100 species are needed to accurately reproduce the ionization fraction. Our preliminary
studies find similar results and we leave for future work for more comprehensive
studies of this region.

\section{Summary and Discussion}\label{sec:con}

In this paper, we aim to construct reduced chemical reaction networks for PPDs that both reproduce the abundances of key chemical species of interest, and is simple enough to use time-dependently in hydrodynamic/magnetohydrodynamic simulations of PPD gas dynamics. In this context, the chemical species of prime interest are those related to disk ionization (electrons and major ions), thermodynamics (important coolants) and abundant species that affect planet formation (e.g., CO, H$_2$O). The species based method can also be applied to other species of interest, and can help identify important chemical processes associated with those species.
Starting from a complex chemical reaction network with 450 gas-phase species and roughly 4000 gas-phase reactions, we follow species-based network reduction method and conduct network reduction at both the midplane and disk surface regions (with photoreactions) over a wide range of radii.

In the disk midplane region, we find that around 28 gas phase species ($\sim$95\% reduction) are adequate to reproduce the ionization fraction and the abundances of major species such as CO and H$_2$O over a wide range of radii. Relative error in ionization fraction is well within a factor of two up to two scale heights above or below disk midplane. Dominant ions and neutral species are also well captured by the reduced network in most disk regions. In the disk surface above the FUV front, roughly 20 gas phase species are found to be sufficient to reproduce the ionization fraction as well as the abundances of major coolants. 

However, right below the FUV front (i.e., intermediate layer in between the midplane and surface), the chemistry is much more complex, and we find $\sim$ 80 species are needed to adequately reproduce the bulk chemistry in this region.  Furthermore this underestimates the complexity since grain surface chemistry is not considered in this study, and they certainly play an important role at intermediate disk heights. Because of this missing link, our reduced network is not yet appropriate in dealing with all problems regarding coupling chemistry with dynamics (e.g., turbulent mixing of radicals from surface towards the midplane \citep{SemenovWiebe11}).

We have ignored grain surface reactions other than H$_2$ formation in this work. Non-thermal desorption and self shielding are considered but are treated
in a simplified fashion. The natural next step is to study the application of this method to grain-surface chemistry. In the mean time, grain-surface reactions also important in the intermediate layer that we have not covered in this work. This is again left for future studies which will be essential for applications requiring chemistry information across the entire disk vertical extent. 

Overall, our results have demonstrated that small scale chemical networks can reasonably reproduce important chemistry in both the midplane and surface layers of PPDs. They can potentially be implemented to be coupled with hydrodynamics/magnetohydrodynamics at affordable computational cost, allowing for more realistic and comprehensive studies of PPDs that couple chemistry with gas dynamics.

\acknowledgments
We are indebted to an anonymous referee whose carefulness and rigor leads to significant improvements of our manuscript. We also thank D. Semenov, L.I. Cleeves and Fujun Du for useful discussions. XNB acknowledges support from Tsinghua University.

\appendix

\section{A. Time dependent errors caused by reduced network}\label{appen:A}

In this Appendix, we discuss the errors in the reduced network by comparing the reduced and full networks throughout the duration of chemical evolution. This may be considered as a first step to assess the robustness of the reduced network when coupling to dynamics, as fluid advection constantly stirs chemistry out of equilibrium. While we are not doing network reduction at a series of times (to account for different chemical evolutionary stages) and combining the results together, we find that fortunately, the reduced network constructed at the final time { ($t=10^6$ yr)} works equally well at earlier times (see below). This result has already being discussed in \citet{Wiebe03}.

\begin{figure}[ht]
\centering
\includegraphics[width =0.3\textwidth]{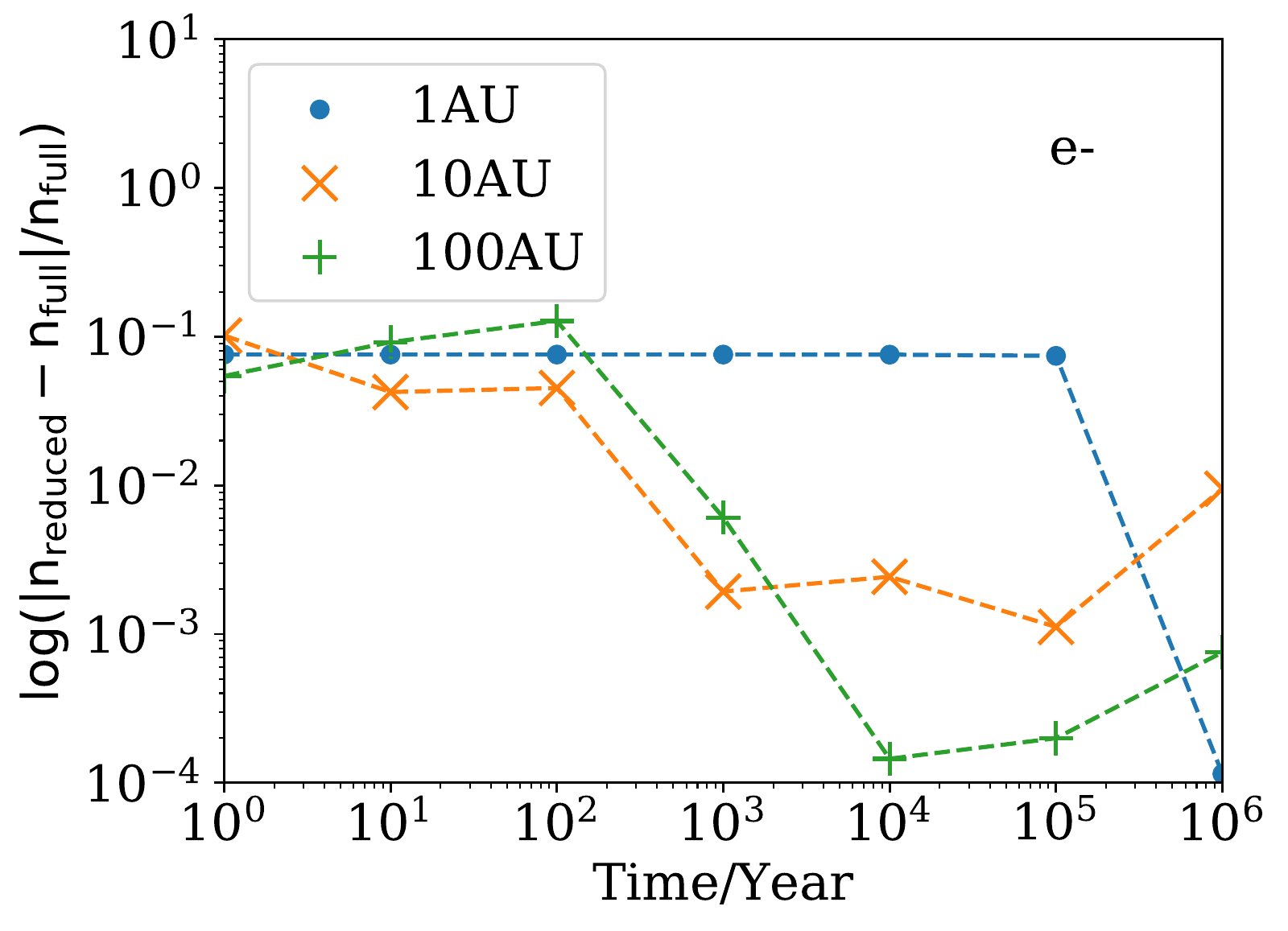} 
\includegraphics[width =0.3\textwidth]{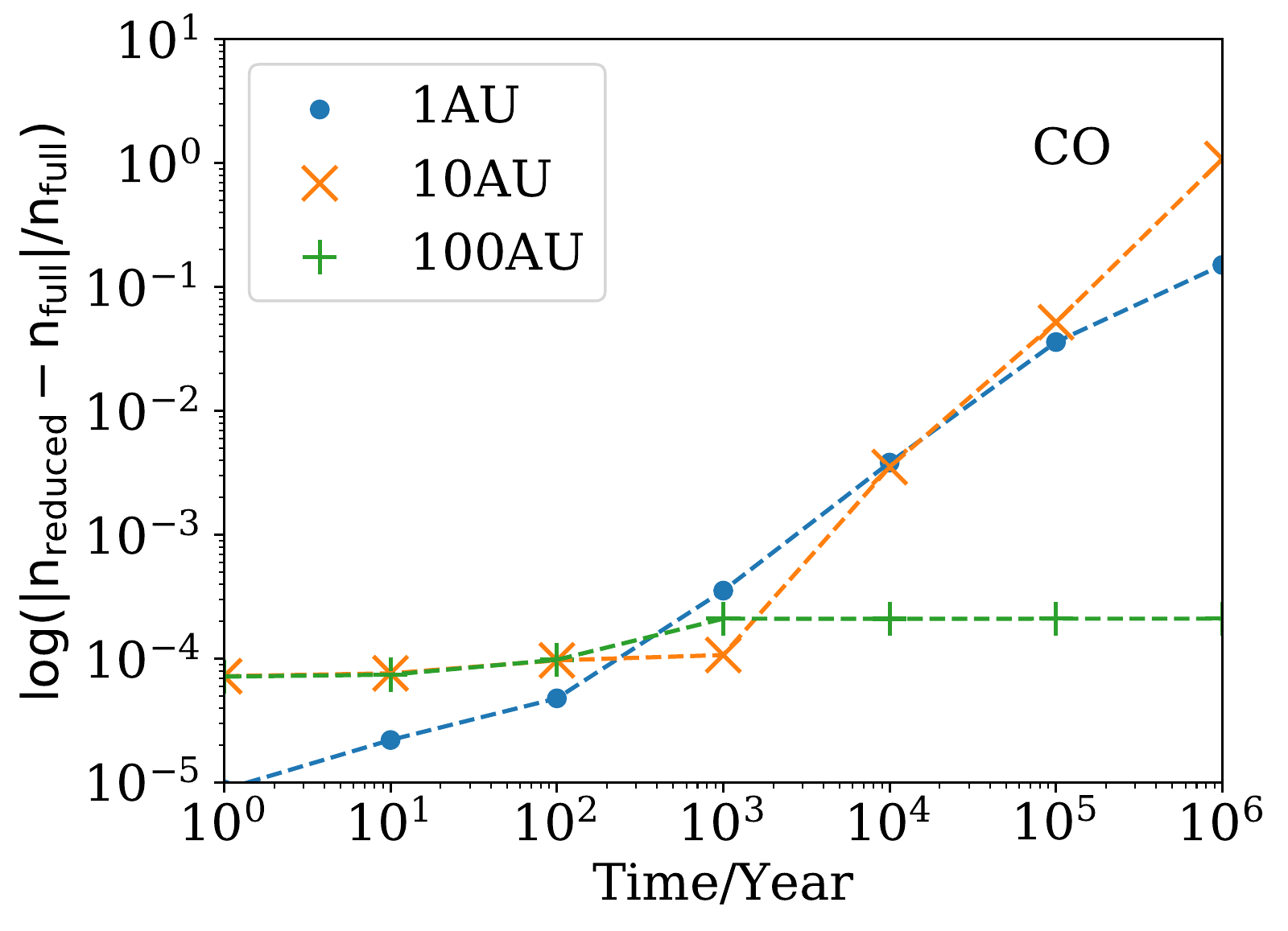} 
\includegraphics[width =0.3\textwidth]{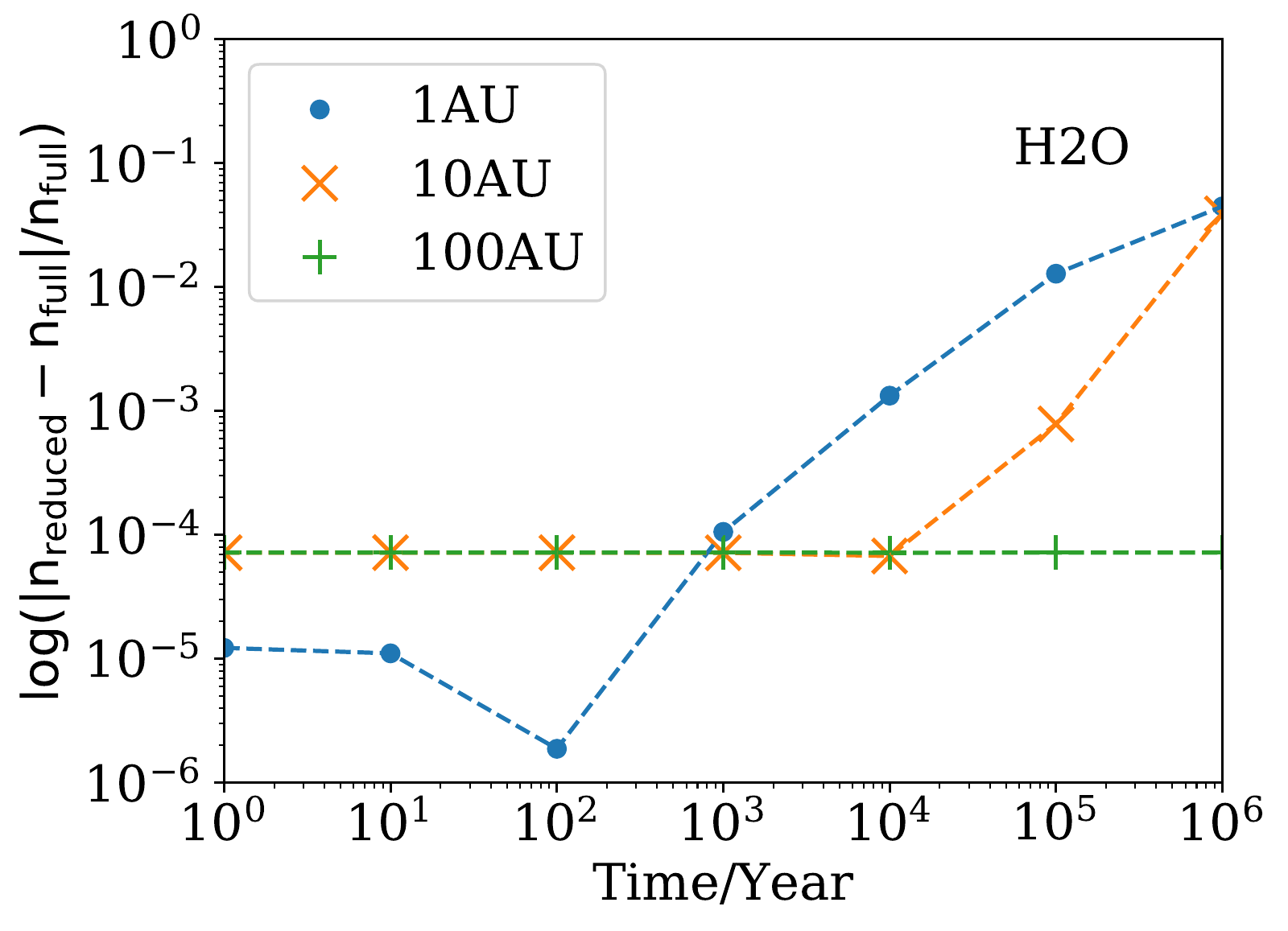} 
\includegraphics[width =0.3\textwidth]{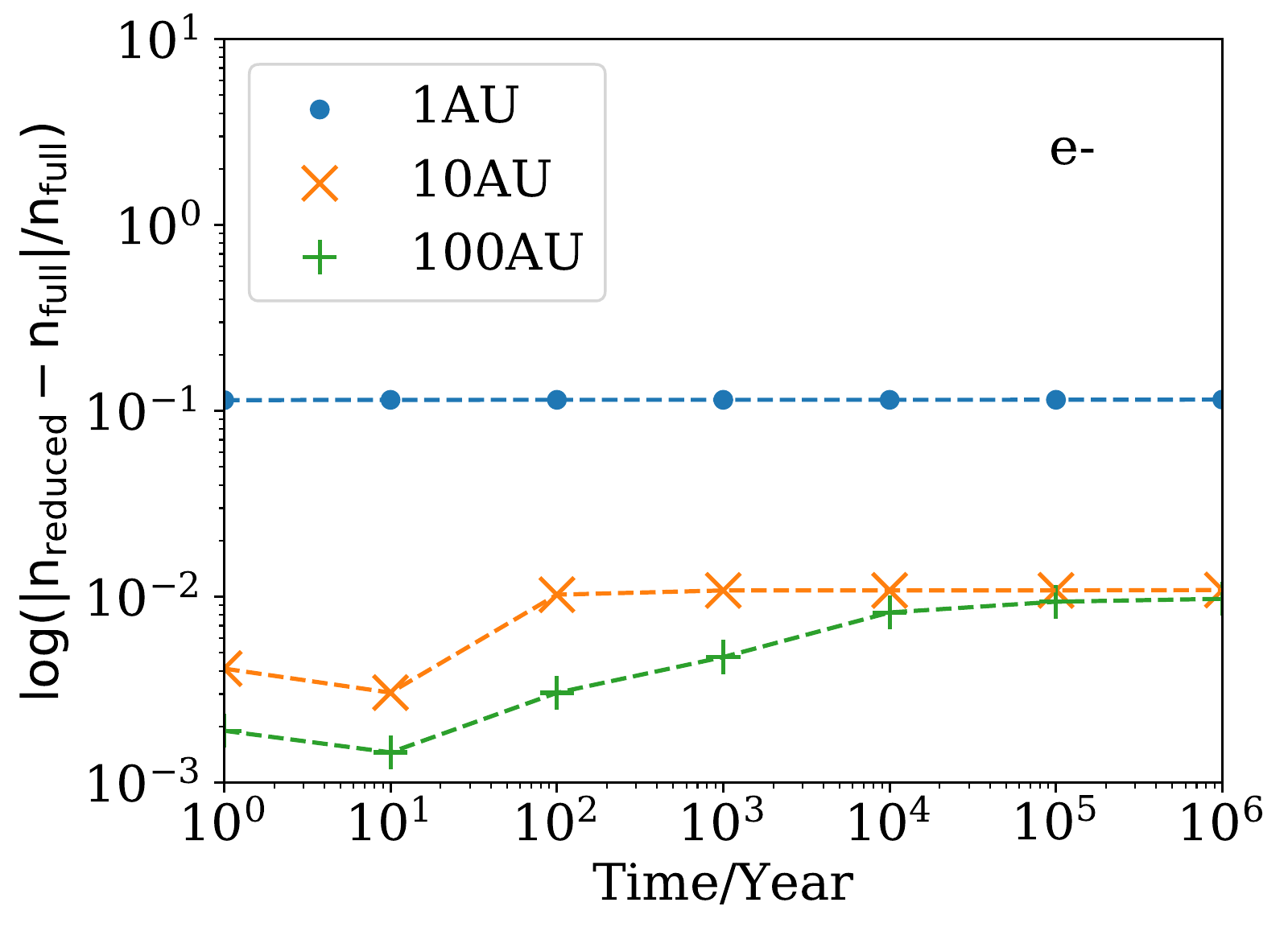} 
\includegraphics[width =0.3\textwidth]{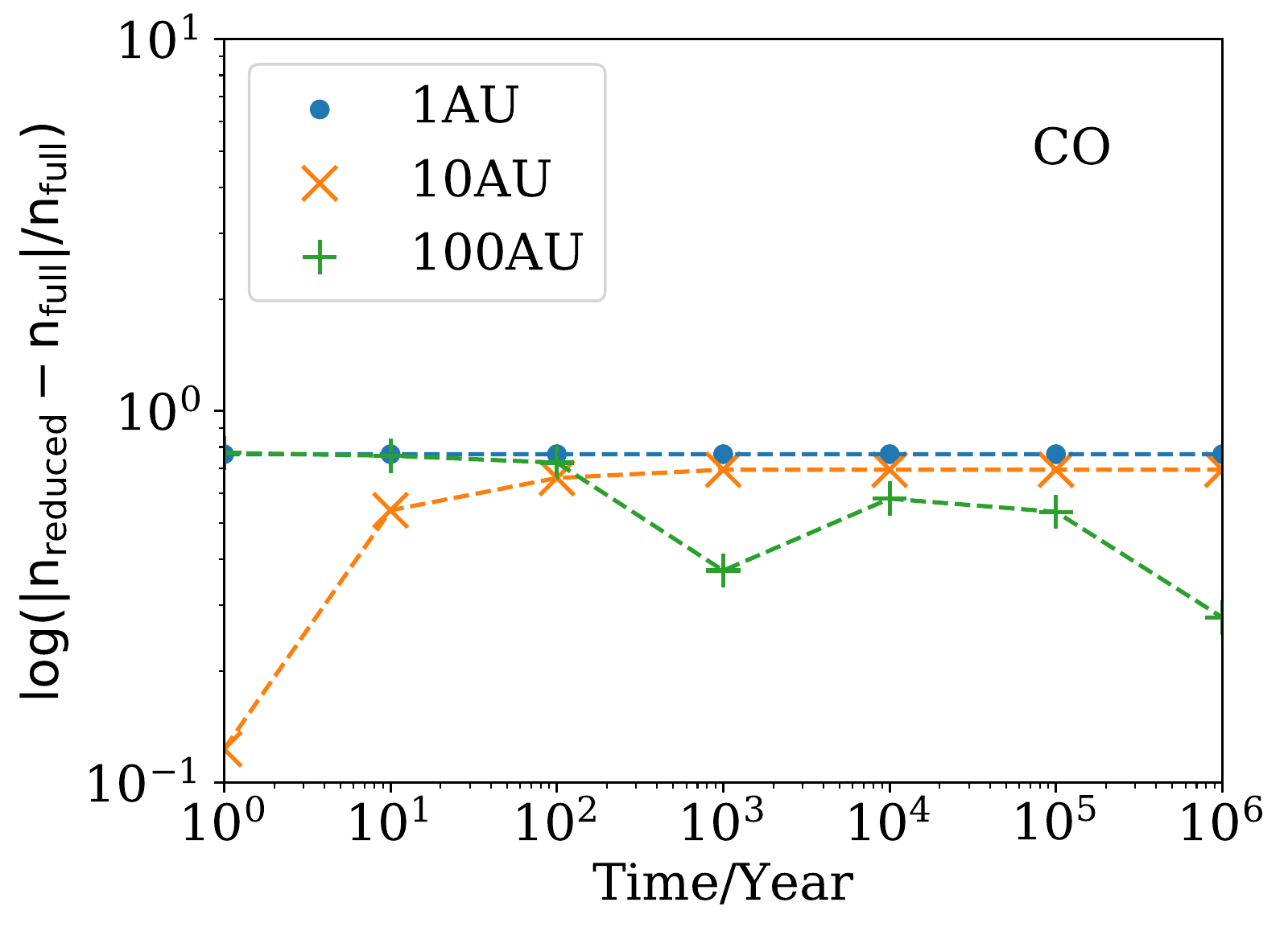}
\includegraphics[width =0.3\textwidth]{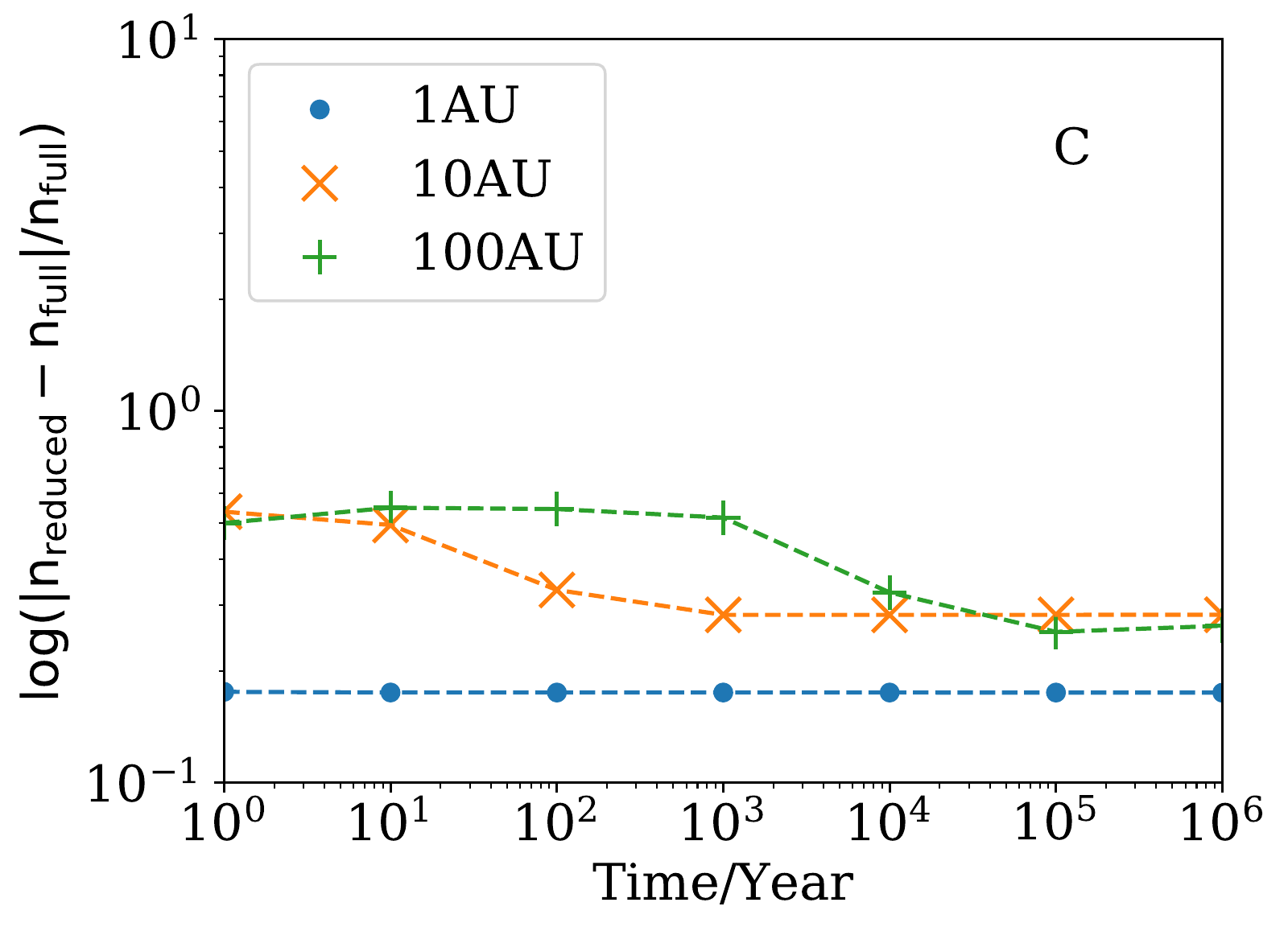}

\label{fig:error}
\caption{Time dependence of relative errors of various major species $\log \left(|n_{\rm reduced}-n_{\rm full}|/n_{\rm full}\right)$, computed at disk midplane (top panels) and disk surface where $\Sigma = 10^{-3}\rm g\cdot cm^{-2}$ (bottom panels).}
\end{figure}

Figure \ref{fig:error} shows the time evolution of the relative errors $\log\left(|n_{\rm reduced}-n_{\rm full}|/n_{\rm full}\right)$ for a number of major species between our reduced network and the full network. We see that for both the disk midplane (top panels) and disk surface (bottom panels) regions, the relative error is smaller than $\sim$ 10\% for majority of species (except CO in the disk midplane at 10AU) during all evolution times. While this is encouraging, we comment that further tests (e.g., in 1D) are needed to fully assess the robustness of the reduced network in a dynamical model.

\section{B. Reaction lists in disk midplane and in the disk surface}\label{appen:reactionlist}

In the disk midplane, the combined reduced network for 1AU, 10AU and 100AU contains 28 gas phase species and 53 gas phase reactions. In the disk surface, reduced network contains 23 gas phase species and 58 gas phase reactions.

\begin{table}
\begin{center}\caption{Gas phase reaction list at the disk midplane\label{tab:mid_reaction}}
\begin{tabular}{l|lll}
\tableline

\multirow{3}*{Charge exchange reactions } & H$^+$ +  CH$_4$ $\rightarrow$ CH$_4^+$ + H & 
H$^+$ + H$_2$O $\rightarrow$ H$_2$O$^+$ + H & He$^+$ + N$_2$ $\rightarrow$ N$_2^+$ + He  \\
 & H$_2^+$ + CH$_4$ $\rightarrow$ CH$_4^+$ + H$_2$ & 
H$_2^+$ + CO  $\rightarrow$ CO$^+$ + H2 &   
H$_2$ + He$^+$ $\rightarrow$ He + H$_2^+$ \\
& He$^+$ + CH$_4$ $\rightarrow$ CH$_4^+$ + He  & 
He$^+$ + H$_2$O $\rightarrow$ H$_2$O$^+$  + He  & 
\\
\hline

\multirow{3}*{Dissociative recombinations} & CH$_5^+$ + e$^-$  $\rightarrow$ CH$_4$ + H & 
H$_3^+$ + e$^-$  $\rightarrow$ H$_2$ + H & 
H$_3^+$ + e$^-$ $\rightarrow$ H + H + H \\
  & H$_3$O$^+$ + e$^-$ $\rightarrow$  H$_2$O + H & 
HCO$^+$ + e$^-$ $\rightarrow$  CO + H & 
N$_2$H$^+$ + e$^-$ $\rightarrow$  N$_2$ +  H \\
& NH$_4^+$ + e$^-$ $\rightarrow$  NH$_3$ + H & & \\
\hline

\multirow{2}*{CRs and X-ray induced reactions}  & H$_2$ $\rightarrow$ H$^+$ + H + e$^-$ (0.021) &    
  H$_2$ $\rightarrow$ H$_2^+$ + e$^-$ (0.883)   &
  H$_2$ $\rightarrow$ H + H (0.096) \\ 
 & He $\rightarrow$ He$^+$ + e$^-$ (0.478) &
  H  $\rightarrow$  H$^+$ + e$^-$ (0.440) & \\   

\hline
Neutral-neutral reactions & H + H $\rightarrow$ H$_2$ (grain surface) &  & \\
\hline

\multirow{10}*{Ion-neutral reactions} & C$^+$ +  H$_2$O $\rightarrow$ HCO$^+$ +  H & 
CH$_4$ + N$_2$H$^+$ $\rightarrow$  N$_2$ +  CH$_5^+$ & 
CH$_5^+$ +  CO $\rightarrow$ HCO$^+$ + CH$_4$ \\
& CH$_5^+$ + H$_2$O $\rightarrow$ H$_3$O$^+$ +  CH$_4$ & 
CO + N$_2$H$^+$ $\rightarrow$  HCO$^+$ +  N$_2$ & 
H$^+$ + CH$_4$ $\rightarrow$ CH$_3^+$ + H$_2$ \\
& H$_2^+$ + H$_2$ $\rightarrow$  H$_3^+$ + H & 
H$_2$ + CH$_2^+$ $\rightarrow$  CH$_3^+$ + H & 
H$_2$ + CH$_4^+$ $\rightarrow$  CH$_5^+$ +  H \\
& H$_2$ + CO$^+$ $\rightarrow$ HCO$^+$ +  H & 
H$_2$ + H$_2$O$^+$ $\rightarrow$  H$_3$O$^+$ + H & 
H$_2$ + He$^+$ $\rightarrow$ He + H$^+$ + H \\
& H$_2$ + N$_2^+$ $\rightarrow$ N$_2$H$^+$ + H & 
H$_2$ + OH$^+$ $\rightarrow$ H$_2$O$^+$ + H &  
H$_2$O + HCO$^+$ $\rightarrow$  CO + H$_3$O$^+$ \\
& H$_2$O + N$_2$H$^+$ $\rightarrow$  N$_2$ +  H$_3$O$^+$ & 
H$_3^+$ + CH$_4$ $\rightarrow$ CH$_5^+$ + H$_2$ & 
H$_3^+$ + CO $\rightarrow$  HCO$^+$ + H$_2$  \\
& H$_3^+$ + H$_2$O $\rightarrow$ H$_3$O$^+$ + H$_2$ & 
H$_3^+$ + N$_2$ $\rightarrow$  N$_2$H$^+$ +  H$_2$ & 
H$_3^+$ + NH$_3$ $\rightarrow$ NH$_4^+$ + H$_2$\\
& H + CH$_4^+$ $\rightarrow$  CH$_3^+$ +  H$_2$  &
H + CH$_5^+$ $\rightarrow$  CH$_4^+$ +  H$_2$  & 
He$^+$ + CH$_4$ $\rightarrow$ CH$_2^+$ + He + H$_2$  \\
& He$^+$ + CH$_4$ $\rightarrow$ CH$_3^+$ +  He + H &
He$^+$ + CO$_2$ $\rightarrow$ O$_2$ +  C$^+$ + He & 
He$^+$ H$_2$O $\rightarrow$ OH$^+$ + He + H \\
& NH$_3$ + H$_3$O$^+$ $\rightarrow$  NH$_4^+$ +  H$_2$O & 
C$^+$ +  CO$_2$ $\rightarrow$ CO$^+$ + CO & NH$_3$ + HCO$^+$ $\rightarrow$  CO + NH$_4^+$ \\
\hline
Radiative associations &  H$_2$ + CH$_3^+$ $\rightarrow$  CH$_5^+$  & 
H$_2$ + C$^+$ $\rightarrow$  CH$_2^+$  & \\
\tableline
\end{tabular}
\end{center}
For CRs and X-ray induced reactions, the number in the parenthesis indicates the coefficient (branching ratio) relative to $\zeta_{\rm eff}$ (total ionization rate on H$_2$). H$_2$ formation on grain surface is treated using Equation (\ref{eq:h2}).
Rate coefficients for all other reactions are chosen according to the UMIST database. 
\end{table}

\begin{table}
\begin{center}
\caption{Gas phase reaction list at the disk surface\label{tab:surface_reaction}}
\begin{tabular}{l| lll}
\tableline

\multirow{3}*{Charge exchange reactions} & S + C$^+$ $\rightarrow$  C + S$^+$  & C$^+$ + Si $\rightarrow$ Si$^+$ + C &
H$^+$ + H2O $\rightarrow$ H$_2$O$^+$ +  H \\
  & H$^+$ + O $\rightarrow$ O$^+$ +  H &
H$^+$ + OH $\rightarrow$  OH$^+$ + H &
H$^+$ +  S $\rightarrow$ S$^+$ +  H\\
 & H + CO$^+$ $\rightarrow$ CO + H$^+$ &
H + H$_2^+$ $\rightarrow$ H$_2$ +  H$^+$ &
H + O$^+$ $\rightarrow$  O + H$^+$\\

\hline

\multirow{5}*{Dissociative recombinations} & CH$^+$ + e$^-$ $\rightarrow$  C + H 
& CO$^+$ + e$^-$ $\rightarrow$  O + C&
H$_2^+$ + e$^-$ $\rightarrow$  H + H \\

& H$_2$O$^+$ + e$^-$ $\rightarrow$  O + H$_2$ 
& H$_2$O$^+$ + e$^-$ $\rightarrow$  O + H + H &
H$_2$O$^+$ + e$^-$ $\rightarrow$  OH + H \\

 & H$_3$O$^+$ + e$^-$ $\rightarrow$  H$_2$O + H
& H$_3$O$^+$ + e$^-$ $\rightarrow$  O + H$_2$ + H &
H$_3$O$^+$ + e$^-$ $\rightarrow$  OH + H$_2$ \\

& H$_3$O$^+$ +  e$^-$ $\rightarrow$  OH + H + H
& HCO$^+$ + e$^-$ $\rightarrow$  CO + H&
OH$^+$ + e$^-$ + O + H \\

& SiO$^+$ + e$^-$ $\rightarrow$  Si + O & & \\

\hline

\multirow{2}*{CRs and X-ray induced reactions} & H$_2$ $\rightarrow$ H$^+$ + H + e$^-$ (0.021)&   
H$_2$  $\rightarrow$ H$_2^+$ + e$^-$  (0.883) &
H$_2$  $\rightarrow$ H + H (0.096) \\
& H  $\rightarrow$  H$^+$ + e$^-$  (0.440) & & \\

\hline

\multirow{2}*{Neutral-neutral reactions} &  
H + H $\rightarrow$ H$_2$ (grain surface) & 
C + OH $\rightarrow$ CO + H &
H$_2$ + O $\rightarrow$ OH + H\\
 & H$_2$ + OH $\rightarrow$ H$_2$O + H &
H + H$_2$O $\rightarrow$ OH + H$_2$ &
H + OH $\rightarrow$ O + H$_2$\\

\hline

\multirow{4}*{Ion-neutral reactions} & C$^+$ + H$_2$O + HCO$^+$ +  H &
C$^+$ + OH $\rightarrow$  CO$^+$ + H&
H$_2^+$ + O $\rightarrow$ OH$^+$ + H\\
 & H$_2$ + C$^+$ $\rightarrow$  CH$^+$ + H &
H$_2$ + CO$^+$ $\rightarrow$ HCO$^+$ +  H&
H$_2$ + H$_2$O$^+$ $\rightarrow$  H$_3$O$^+$ + H \\
& H$_2$ + O$^+$ $\rightarrow$  OH$^+$ + H &
H$_2$ + OH$^+$ $\rightarrow$ H$_2$O$^+$ + H &
H + CH$^+$ $\rightarrow$ C$^+$ +  H$_2$ \\
& OH + Si$^+$ $\rightarrow$ SiO$^+$ +  H & & \\
\hline

\multirow{2}*{Radiative associations} & H$^+$ +  H $\rightarrow$ H$_2^+$ &
H + C$^+$ $\rightarrow$  CH$^+$ &
H + O $\rightarrow$ OH \\
& H + OH  $\rightarrow$ H$_2$O &  &\\

\hline

\multirow{2}*{Radiative recombinations} & C$^+$ + e$^-$ $\rightarrow$  C &
H$^+$ + e$^-$ $\rightarrow$  H  & 
S + e$^-$  $\rightarrow$ S \\
& Si$^+$ + e$^-$ $\rightarrow$  Si & & \\

\hline

\multirow{3}*{\bf Photo reactions} & CO $\rightarrow$ C + O &
H$_2$O $\rightarrow$ H$_2$O$^+$ + e$^-$ &
H$_2$O $\rightarrow$ OH + H \\
& OH $\rightarrow$ O + H &
OH $\rightarrow$ OH$^+$ + e$^-$  &
S $\rightarrow$ S$^+$ + e$^-$  \\
& Si  $\rightarrow$ Si$^+$ +  e$^-$  &   C $  \rightarrow$ C$^+$ + e$^-$  & \\

\tableline
\end{tabular}
\end{center}
For CRs and X-ray induced reactions, the number in the parenthesis indicates the coefficient (branching ratio) relative to $\zeta_{\rm eff}$ (total ionization rate on H$_2$). H$_2$ formation on grain surface is treated using Equation (\ref{eq:h2}).
Rate coefficients for all other reactions are chosen according to the UMIST database. 
\end{table}

\newpage
\bibliographystyle{apj}
\bibliography{chem}
\label{lastpage}
\end{document}